\documentclass{article}

\usepackage{amsfonts,epsfig,fullpage,amsmath,psfig,color}

\begin{document}

\newcommand{\xx}{{\mathbf x}}
\newcommand{\vv}{{\mathbf v}}
\newcommand{\uu}{{\mathbf u}}

\newcommand{\ee}{{\mathrm e}}

\newcommand{\one}{\mathbf{1}}
\newcommand{\ones}{\mathbf{1}}

\newcommand{\C}{{\cal C}}
\def\thefootnote{\fnsymbol{footnote}}

\newcommand{\ra}{\rightarrow}

\newcommand{\e}{\epsilon}
\newcommand{\dd}{\delta}

\def\limninf{\lim_{n \rightarrow \infty}}

\newtheorem{definition}{Definition}
\newtheorem{lemma}{Lemma}
\newtheorem{claim}{Claim}
\newtheorem{theorem}{Theorem}
\newtheorem{corollary}{Corollary}
\newtheorem{conjecture}{Conjecture}

\newcommand{\amax}{\alpha_{\max}}
\newcommand{\gmax}{g_{\max}}

\def\a{\alpha}

\def\e{\varepsilon}
\def\CC{{\cal C}}
\def\FF{F}

\renewcommand{\Pr}{{\mathbf {P}}}
\newcommand{\dima}[1]{{{#1}}}

\newcommand{\dimt}[1]{{{#1}}}

\newcommand{\dimss}[1]{{{#1}}}

\newcommand{\dimf}[1]{{{#1}}}

\newcommand{\dims}[1]{{{#1}}}

\def\as{a.s.\ }
\def\whp{w.h.p.}
\def\Whp{W.h.p.}

\def\eee{\end{document}}

\def\eg{e.g.,\ }
\def\ie{i.e.,\ }
\newcommand{\fo}[2]{\ensuremath{{F}_{#1}(n,#2)}}
\def\var{{\rm var}}
\def\ex{{\mathbf E}}
\def\Up{\Upsilon}

\title{The Threshold for Random $k$-SAT is $2^k \log 2- O(k)$}

\author{
Dimitris Achlioptas \\
Microsoft Research, Redmond, WA \\
{\tt optas@microsoft.com} \and Yuval Peres\thanks{Research
supported by NSF Grant DMS-0104073, NSF Grant DMS-0244479 and a
Miller Professorship at UC Berkeley. Part of this work was done
while visiting Microsoft Research.}\\
Department of Statistics, University of California, Berkeley, CA\\
{\tt peres@stat.berkeley.edu} }

\maketitle
\begin{abstract}
\dima{Let $F_k(n,m)$ be a random $k$-CNF formula formed by
selecting uniformly and independently $m$ out of all possible
\mbox{$k$-clauses} on $n$ variables. It is well-known that if $r
\geq 2^k \log 2$, then $F_k(n,rn)$ is unsatisfiable with
probability {that tends to 1 as $n \to \infty$. We prove that
if $r \leq 2^k \log 2 - t_k$, where $t_k = O(k)$, then $F_k(n,rn)$ is satisfiable \mbox{with probability
that tends to 1 as $n \to \infty$.}}

Our technique, in fact, yields an explicit lower bound for the
random $k$-SAT threshold for every $k$. For $k \geq 4$ our bounds
improve all previously known such bounds.
}\end{abstract}

\section{Introduction}

Call a disjunction of $k$ Boolean variables a $k$-clause. For a
set $V$ of $n$ Boolean variables, let $C_k(V)$ denote the set of
all $2^k n^k$ possible $k$-clauses on $V$. A random $k$-CNF
formula $F_k(n,m)$ is formed by selecting uniformly, independently
and with replacement $m$ clauses from $C_k$ and taking their
conjunction\footnotemark[2]. The study of such random $k$-CNF
formulas has attracted substantial interest in logic,
optimization, combinatorics, theory of algorithms and, more
recently, statistical physics.

Say that a sequence of events ${\mathcal E}_n$ occurs with high
probability (w.h.p.) if $\lim_{n \ra \infty} \Pr[{\mathcal
E}_n]=1$ and with uniformly positive probability if $\liminf_{n
\ra \infty} \Pr[{\mathcal E}_n]>0$. We emphasize that throughout
the paper $k$ is arbitrarily large but fixed, while $n \rightarrow
\infty$. For each $k \geq 2$, let
\begin{eqnarray*}
 r_k   & \equiv & \sup\{r : F_k(n,rn)\mbox{ is satisfiable    \whp}\} \enspace ,\\
 r_k^* & \equiv & \inf\{r : F_k(n,rn)\mbox{ is unsatisfiable  \whp}\}
\enspace .
\end{eqnarray*}

\footnotetext[2]{Our results hold in all common models for random
$k$-SAT, \eg when clause replacement is not allowed.
See Section~\ref{ground}.}

Clearly, $r_k \leq r_k^*$. The {\em Satisfiability Threshold
Conjecture\/} asserts that $r_k=r_k^*$ for all $k \ge 3$. Our main
result establishes an asymptotic form of this conjecture.
\begin{theorem}\label{bain}
As $k \to \infty$,
\[
r_k=r_k^*(1-o(1)) \enspace .
\]
\end{theorem}

\dimf{As we will see in Section~\ref{baka},} a classical and very
simple argument gives $r_k^* \leq 2^k \log 2$. The following
theorem implies that this bound is asymptotically tight. The
theorem also sharpens the $o(1)$ term in Theorem~\ref{bain}.
\begin{theorem}\label{main} There exists a sequence $\delta_k
\rightarrow 0$ such that for all $k \geq 3$,
$$
r_k \geq 2^k \log 2 - (k+1) \frac{\log 2}{2} - 1 - \delta_k
\enspace .
$$
\end{theorem}

Theorem~\ref{main} establishes that $r_k\sim 2^k \log 2$, in
agreement with the predictions of Monasson and
Zecchina~\cite{MZRK} based on the ``replica method" of statistical
mechanics. Like most arguments based on the replica method, the
\dima{approach} in~\cite{MZRK} is mathematically sophisticated
\dima{but far from rigorous. To the best of our knowledge,} our
result \dimf{is} the first rigorous proof of a replica method
prediction for \dimf{any} \mbox{NP-complete} problem at zero
temperature.\medskip

Obtaining tight bounds for the \dimf{thresholds $r_k,r_k^*$} is a
{benchmark problem for a number of analytic and combinatorial
techniques of wider applicability}~\cite{LLL,frie,JSV,mick}. The
best bounds prior to our work for general $k$, from~\cite{naesat}
and~\cite{DuBo} respectively, differed \dima{roughly by a factor
of 2:}
$$
2^{k-1} \log 2 - \Theta(1) \le r_k \le r_k^* \le  2^k \log 2 -
\Theta(1) \enspace . \qquad
$$

Traditionally, lower bounds for $r_k$ have been established by
analyzing algorithms for finding satisfying assignments, \dimf{\ie
by proving in each case that some specific algorithm succeeds
\whp\ on $F_k(n,rn)$ for $r$ smaller than a certain value}.
Indeed, until very recently, all lower bounds for $r_k$ were
algorithmic and of the form $\Omega(2^k/k)$. The bound $r_k \geq
2^{k-1} \log 2 -\Theta(1)$ from~\cite{naesat}, derived via a
non-algorithmic argument, was the first to break the $2^k/k$
barrier.

Our proof of Theorem~\ref{main} is also non-algorithmic, based
instead on a delicate application of the second moment method. By
not going after some particular satisfying truth assignment, as
algorithms do, our arguments offer some glimpses of the
``geometry'' of the set of satisfying truth assignments. {Also,}
the proof yields an explicit lower bound for $r_k$ for each $k
\geq 3$. Already for $k \geq 4$ this improves all previously known
lower bounds for $r_k$. Below, we compare our lower bound with the
best known algorithmic lower bound~\cite{FrSu,342} and the best
known upper bound~\cite{dub_announ,DuBo,KKKS} for some small
values of $k$.

\begin{table*}[h]\label{tab:val}
\centering $
\begin{array}{c|ccccccc}
 k                              &  3        & 4         & 5         & 7         & 10        & 20        & 21 \\   \hline
\mbox{Upper bound}              & 4.51      & 10.23     & 21.33     & 87.88     & 708.94    & 726,817   & 1,453,635 \\
\mbox{Our lower bound}          & 2.68      & 7.91      & 18.79     & 84.82     & 704.94    & 726,809   & 1,453,626 \\
\mbox{Algorithmic lower bound}  & 3.42      & 5.54      & 9.63      & 33.23     & 172.65    & 95,263    & 181,453 \\
\end{array}
$
\end{table*}

\subsection{Background}\label{baka}

Franco and Paull~\cite{FrPa}, in the early 80's, observed that
$r_k^* \le 2^k \log 2$. To see this, fix any truth assignment and
observe that a random $k$-clause is satisfied by it with
probability $1-2^{-k}$. Therefore, the expected number of
satisfying truth assignments of $F_k(n,rn)$ is
$[2(1-2^{-k})^r]^n=o(1)$ for $r \geq 2^k \log 2$. In 1990, Chao
and Franco~\cite{ChFrUC} complemented this by proving that for $r
< 2^k/k$ a simple algorithm, called {\sc Unit Clause (uc)}, finds
a satisfying truth assignment with uniformly positive probability.

At around the same time, experimental results by Cheeseman,
Kanefsky and Taylor~\cite{cheese} and Mitchell, Selman and
Levesque~\cite{MSL} suggested that random $k$-SAT, while a logical
model, also behaves like a physical system in the sense that it
appears to undergo a phase transition. Perhaps the first statement
of the satisfiability threshold conjecture appeared \dimf{about
ten years ago} in the work of Chv\'{a}tal and Reed~\cite{mick} who
proved $r_2 = r_2^* = 1$ and, by analyzing  an extension of {\sc
uc}, established that $r_k \geq (3/8) 2^k/k$. A few years later,
Frieze and Suen~\cite{FrSu} improved this lower bound to $r_k \geq
c_k 2^k/k$, where $\lim_{k \ra \infty} c_k = 1.817\ldots$,
\dimf{and this remained the best bound for $r_k$ until recently.}

\dimf{In a breakthrough paper, Friedgut~\cite{frie} proved the
existence of a {\em non-uniform\/} threshold.}
\begin{theorem}[Friedgut \cite{frie}]
\label{thm:frie} For each $k \geq 2$, there exists a sequence
$r_k(n)$ such that for every $\epsilon>0$,
\[
\limninf \Pr[F_k(n,rn)\mbox{ is satisfiable}] =
\begin{cases}
1 & \mbox{if $r = (1-\epsilon)\, r_k(n)$}\\
0 & \mbox{if $r = (1+\epsilon)\, r_k(n)$} \enspace .
\end{cases}
\]
\end{theorem}

\dima{As mentioned earlier, in~\cite{naesat}, Moore and the first
author established  $r_k \geq 2^{k-1} \log 2 - 1$. Independently,
Frieze and Wormald~\cite{friezewormald} proved that if $k$ is
allowed to grow with $n$, in particular if $k-\log_2 n \rightarrow
+\infty$, then random $k$-SAT has a sharp threshold around
$m=n(2^k+O(1))\log 2$.} See~\cite{naesat} for further background.

The rest of the paper is organized as follows. In the next section
we recall the argument in~\cite{naesat}, highlight its main
weakness and discuss how we overcome it. Our main idea can be
implemented either by a simple weighting scheme or by a more
refined large deviations argument. Both approaches yield $2^k \log
2$ as the leading term in the lower bound for $r_k$. The weighting
scheme argument is more compact and technically simpler. However,
it gives away a factor of four in the $\Theta(k)$ second order
term. The large deviations analysis, on the other hand, is tight
for our \dima{method, up to an additive $O(1)$}. We present the
weighting scheme argument in
Sections~\ref{ground}---\ref{end_weight}. The additional material
for the large deviations argument appears in
Sections~\ref{trunc}---\ref{end_trunc}. In Section~\ref{sec:concrete} we describe our derivation of
explicit lower bounds for small values of $k$. We conclude with
some discussion and open problems in Section~\ref{sec:conc}.

\section{Outline and heuristics}\label{intuit}

\dimt{For any non-negative random variable $X$ one can get a lower
bound on $\Pr[X > 0]$ by the following inequality.
\begin{lemma}\label{lemma:sec}
For any non-negative random variable $X$,
\begin{equation}\label{eq:second}
\Pr[X > 0] \,\ge\, \frac{\ex[X]^2}{\ex[X^2]} \enspace .
\end{equation}
\end{lemma}

In particular, if $X$ denotes the number of satisfying assignments
of a random formula $F_k(n,rn)$, one can get a lower bound on the
probability of satisfiability by applying~\eqref{eq:second} to
$X$. \dima{We will refer to this approach as the ``vanilla"
application of the second moment method.} Indeed, the following
immediate corollary of Theorem~\ref{thm:frie} implies that if
$\Pr[X>0]> 1/C$ for any constant $C>0$, then $r_k \geq r$.
\begin{corollary}\label{cor:boost} Fix $k\geq 2$. If
$F_k(n,rn)$ is satisfiable with uniformly positive probability
then $r_k \geq r$.
\end{corollary}
Thus, if for a given $r$ we have $\ex[X^2] = O(\ex[X]^2)$, then
$r_k \geq r$. Unfortunately, as we will see, this is never the
case: for every constant $r>0$, there exists $\beta= \beta(r)>0$
such that $\ex[X^2] > (1+\beta)^n \, \ex[X]^2$.}

\subsection{The vanilla second moment method fails}

Given a CNF formula ${\FF}$ on $n$ variables let \dims{${\cal
S}({\FF}) = \{\sigma : \sigma \mbox{ satisfies } \FF\} \subset
\{0,1\}^n$} denote the set of satisfying truth assignments of
${\FF}$ and let $X=X(F)=|{\mathcal S}(F)|$. Then, for a $k$-CNF
formula with independent clauses $c_1,c_2,\ldots,c_m$
\begin{equation}\label{eq:smsat}
\ex[X^2] = \ex\left[\left(\sum_{\sigma} \ones_{\sigma \in
{\mathcal S}(F)}\right)^2\right] = \ex\left[\sum_{\sigma,\tau}
\ones_{\sigma,\tau \in {\mathcal S}(F)}\right] =
\sum_{\sigma,\tau}\ex\left[\prod_{c_i}\ones_{\sigma,\tau \in
{\mathcal S}(c_i)} \right] = \sum_{\sigma,\tau}\prod_{c_i}
\ex[\ones_{\sigma,\tau \in {\mathcal S}(c_i)}] \enspace .
\end{equation}
We claim that $\ex[\ones_{\sigma,\tau \in {\mathcal S}(c_i)}]$,
\ie the probability that a fixed pair of truth assignments
$\sigma,\tau$ satisfy the $i$th random clause, depends only on the
number of variables $z$ to which $\sigma$ and $\tau$ assign the
same value. Specifically, if the overlap of $\sigma$ and $\tau$ is
$z= \alpha n$, we claim that this probability is
\begin{equation}\label{fsa}
    \Pr[\sigma,\tau \in {\mathcal{S}}(c_i)]= 1 - 2^{1-k} + 2^{-k}
    \alpha^k \equiv f_S(\alpha)  \enspace .
\end{equation}
This follows by observing that if $c_i$ is not satisfied by
$\sigma$, the only way for it to also not be satisfied by $\tau$
is for all $k$ variables in $c_i$ to lie in the overlap of
$\sigma$ and $\tau$. Thus, $f_S$ quantifies the correlation
between $\sigma$ being satisfying and $\tau$ being satisfying as a
function of their overlap. In particular, observe that truth
assignments with overlap  $n/2$ are uncorrelated since $f_S(1/2) =
(1-2^{-k})^2 = \Pr[\sigma {\mbox{ is satisfying}}]^2$.

Since the number of ordered pairs of assignments with overlap $z$
is $ 2^n \,\binom{n}{z}$ we see that~\eqref{eq:smsat}
and~\eqref{fsa} imply
\[
 \ex[X^2] = 2^n \sum_{z=0}^n  \binom{n}{z} \,f_S(z/n)^{m}
 \enspace .
 \]
Writing $z=\alpha n$ and using  the approximation $\binom{n}{z} =
(\a^\a(1-\a)^{1-\a})^{-n} \times {\mathrm {poly}}(n)$ we get
\[
\ex[X^2]  \geq   2^n \left(\max_{0 \leq \a \leq 1}
            \left[\frac{f_S(\a)^r}{\a^\a(1-\a)^{1-\a}}\right]
        \right)^n \times {\mathrm {poly}}(n)
 \equiv  \left(\max_{0 \leq \a \leq 1} \Lambda_S(\a)\right)^n
\times {\mathrm {poly}}(n) \enspace .
\]

Note now that $\ex[X]^2 = \left(2^n (1-2^{-k})^{rn}\right)^2 =
\left(4f_S(1/2)^r\right)^n=\Lambda_S(1/2)^n$. Therefore, if there
exists some $\a \in [0,1]$ such that
$\Lambda_S(\alpha)>\Lambda_S(1/2)$, then the second moment is
exponentially greater than the square of the expectation and we
only get an exponentially small lower bound for $\Pr[X>0]$. Put
differently, unless the dominant contribution to $\ex[X^2]$ comes
from uncorrelated pairs of satisfying assignments, \ie pairs with
overlap $n/2$, the second moment method fails. Indeed, for any
constant $r>0$ this is precisely what happens as the function
$\Lambda_S$ is maximized at some $\a > 1/2$. The reason for this
is as follows: while the entropic factor {${\mathcal E}(\alpha) =
1/(\a^{\a}(1-\a)^{1-\a})$} is maximized when $\a=1/2$, the
\dims{function $f_S$ has a positive derivative in $(0,1)$.}
Therefore, the derivative of $\Lambda_S$ is never 0 at 1/2,
instead becoming 0 only when the correlation benefit balances with
the penalty of decreasing entropy at some $\a>1/2$.

\vspace*{1cm}

\begin{figure}\label{fig_nae}
\begin{minipage}{2in}
\begin{center}
  \centerline{\psfig{figure=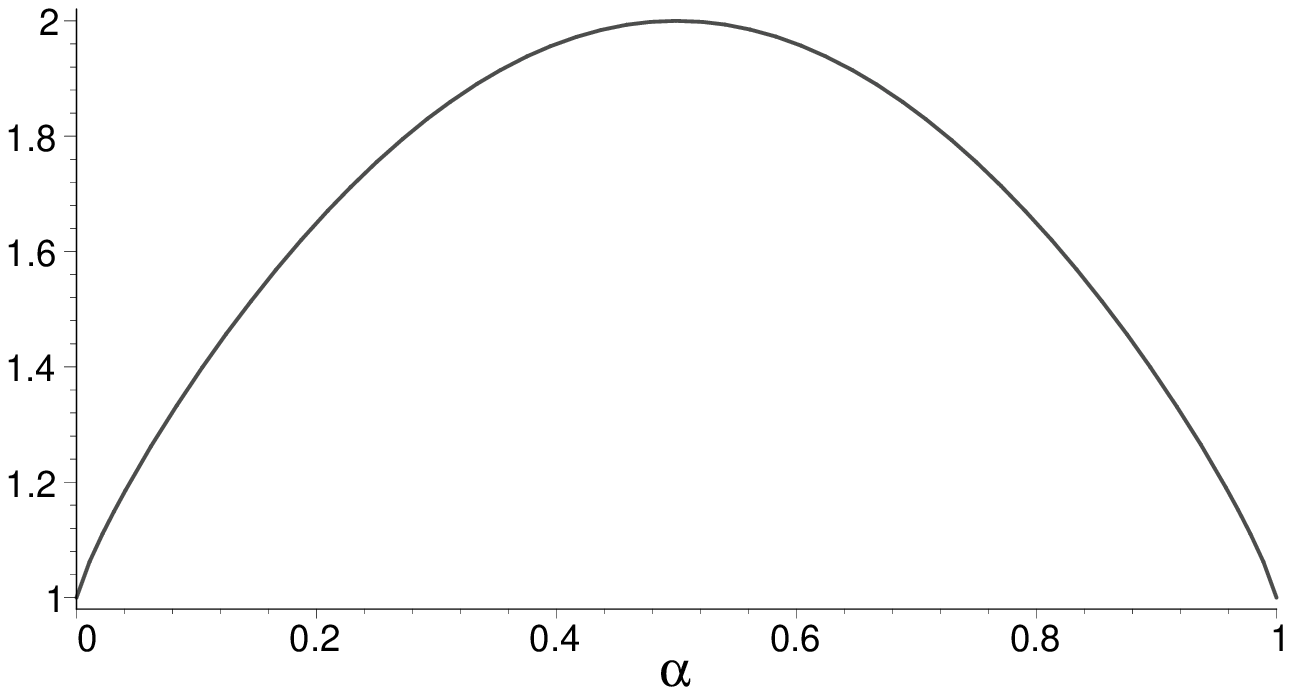,height=1.5in,width=2in}}
  {$\displaystyle{{\mathcal E}(\alpha)={1}/({\alpha^{\alpha}(1-\alpha)^{1-\alpha}}})$}
\end{center}
\end{minipage}
\begin{minipage}{2in}
\begin{center}
  \centerline{\psfig{figure=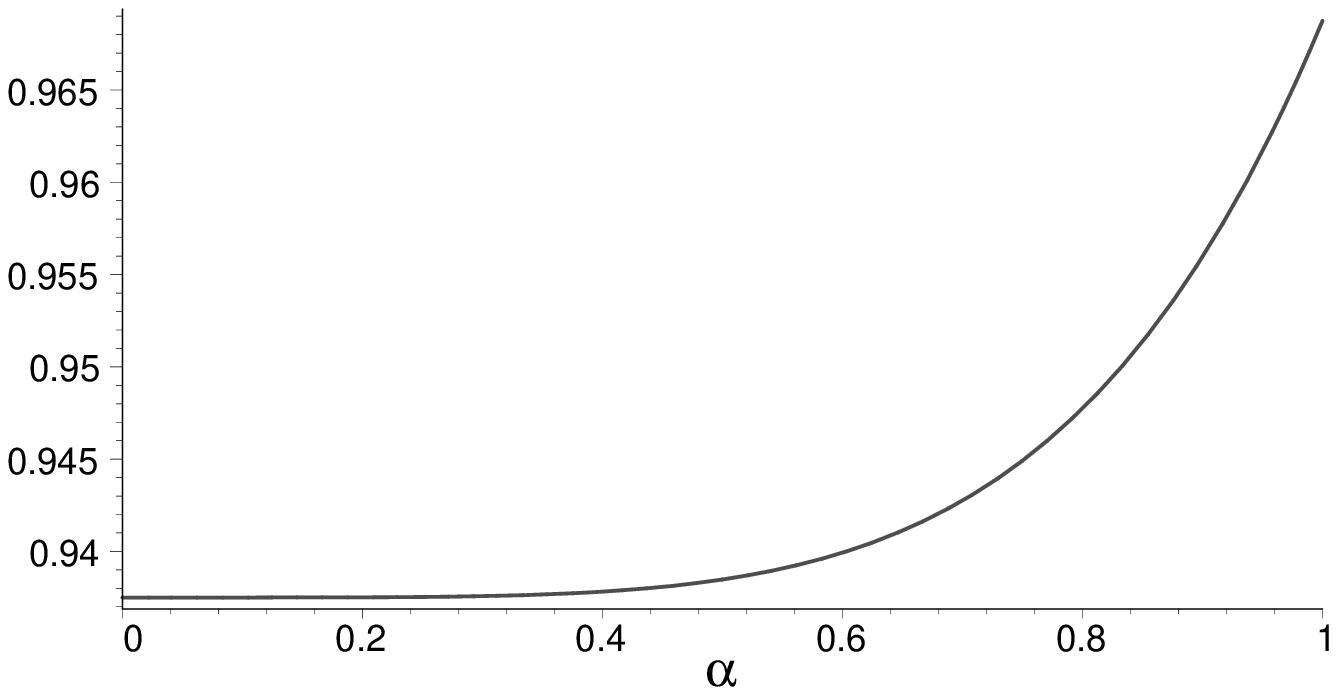,height=1.5in,width=2in}}
  {$f_S(\alpha)=1-2^{1-k}+\alpha^k$}
\end{center}
\end{minipage}
\begin{minipage}{2in}
\begin{center}
  \centerline{\psfig{figure=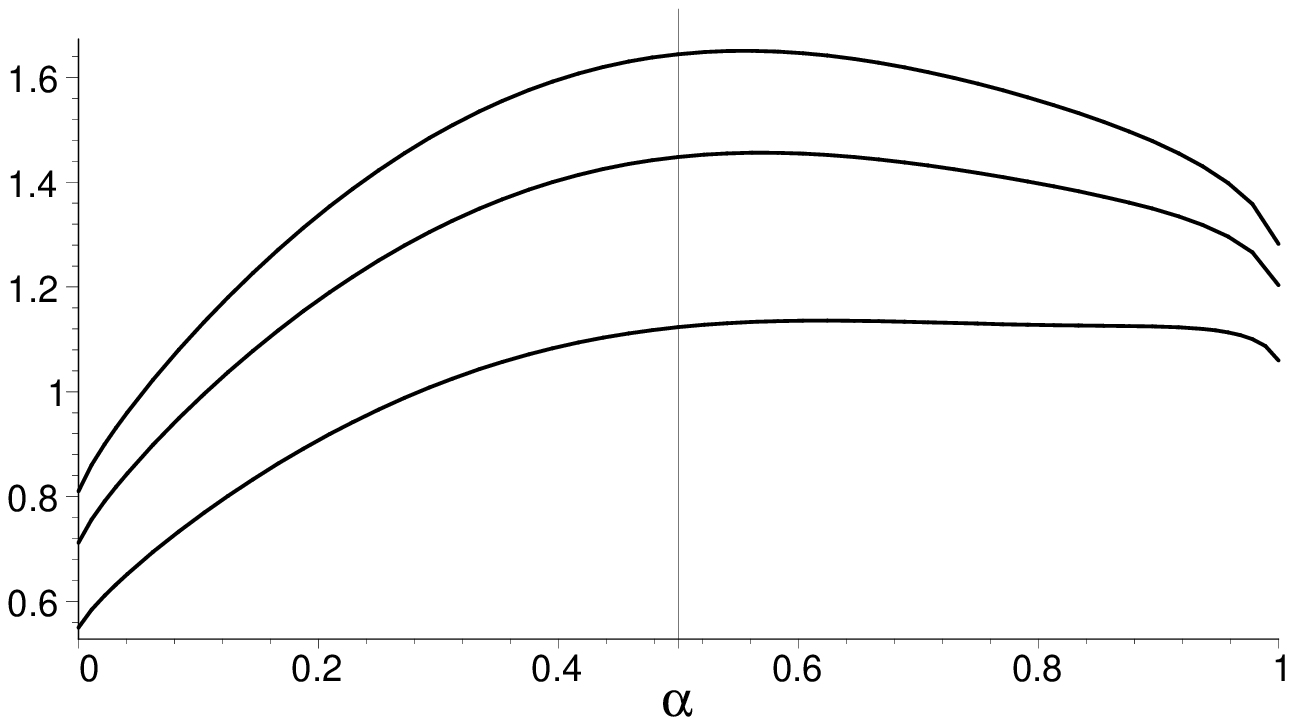,height=1.5in,width=2in}}
{$\Lambda_S(\alpha)=2\mathcal{E}(\alpha)f_S(\alpha)^r$
  }
\end{center}
\end{minipage}
\caption{$k=5$, $r=14,16,20$ (top to bottom).}
\end{figure}

\subsection{Random NAE {\Large \mbox{$k$}}-SAT and balance}

In~\cite{naesat}, the second moment method was applied
successfully by considering only those satisfying truth
assignments whose complement is also satisfying. Observe that this
is equivalent to interpreting $F_k(n,m)$ as an instance of Not All
Equal $k$-SAT, where $\sigma$ is a solution iff under $\sigma$
every clause has at least one satisfied literal {\em and\/} at
least one unsatisfied literal. In particular, if $\sigma,\tau$
have overlap $z=\alpha n$ and $c$ is a random clause
\[
\Pr[\sigma,\tau \mbox{ NAE-satisfy $c$}] = 1 - 2^{2-k} +
2^{1-k}(\alpha^k+(1-\alpha)^k) \equiv f_N(\alpha) \enspace .
\]
The key point is that $f_N$ is symmetric around $\a=1/2$ and, as a
result, the product $\mathcal{E}(\alpha)f_N(\alpha)^r$ always has
a local extremum at $1/2$. In~\cite{naesat} it was shown that for
\mbox{$r \leq 2^{k-1} \log 2-1$ } this extremum is a global
maximum, implying that for such $r$, $F_k(n,m)$ is \whp\
\mbox{[NAE-]} satisfiable. It is worth noting that for $r \geq
2^{k-1} \log 2$, \whp\ $F_k(n,m)$  is {\em not\/} NAE-satisfiable,
\ie the second moment method determines the NAE-satisfiability
threshold within an additive constant. Intuitively, the symmetry
of $f_N$ stems from the fact that NAE-satisfying assignments come
in complementary pairs and, thus, having overlap $z$ with an
NAE-satisfying assignment $\sigma$ (and $n-z$ with
$\overline{\sigma}$) is indistinguishable from having overlap
$n-z$ with $\sigma$ (and $z$ with $\overline{\sigma}$).

{The} suspicion motivating this work is that the correlations
behind the failure of the vanilla second moment method are
{mainly} due to the following form of populism: {\em {satisfying
assignments tend to lean} towards the majority vote truth
assignment.} Observe that truth
assignments that satisfy many literal occurrences in the random formula
have significantly greater probability of being satisfying. At the
same time, such assignments are highly correlated since, in order
to satisfy many literal occurrences, they tend to agree with each
other (and the majority truth assignment) on more than half the
variables.

Note that our suspicion regarding populism is consistent with the
success of the second moment method for random NAE $k$-SAT. In
that problem, since we need to have at least one satisfied {\em
and\/} at least one dissatisfied literal in each clause, leaning
towards majority is a disadvantage. As intuition suggests,
``middle of the road" assignments have the greatest probability of
being NAE-satisfying. Alternatively, observe that conditioning on
$\sigma$ being NAE-satisfying does not increase the expected
number of satisfied literal occurrences under $\sigma$, whereas
conditioning on $\sigma$ being only satisfying increases this
expectation by a factor $2^k/(2^k-1)$ relative to the
unconditional expectation $km/2$. To overcome these correlations,
populism must be discouraged and the delicacy with which this is
done determines the accuracy of the resulting bound.

An example from a different area, which was another inspiration
for our work, is the recent proof of the Erd\H{o}s-Taylor
conjecture from 1960 for the simple random walk in the planar
square lattice (see~\cite{ET},\cite{DPRZ} and for a popular
account~\cite{drunk}). The conjecture was that the number of
visits to the most frequently visited lattice site in the first
$n$ steps of the walk, is asymptotic to $(\log n)^2/\pi$.
Erd\H{o}s and Taylor~\cite{ET} obtained a (sharp) upper bound via
an easy calculation of the expectation of the number $X_a$ of
vertices visited at least $a(\log n)^2$ times. The lower bound
they obtained was four times smaller than the conjectured value.
In that setting the vanilla second moment method fails, since the
events that two vertices $u,v$ are visited frequently are highly
correlated. The conjecture was proved in \cite{DPRZ} by first
recognizing the main source of the correlation in a certain
``populism'' (when the random walk spends a long time in the
smallest disk containing both $u$ and $v$). Replacing $X_a$ by a
weighted count that discourages such loitering, confirmed that
this was indeed the source of excessive correlations as the
weighted second moment was successful.\medskip

In a nutshell, our plan is to apply the second moment method to
{\em balanced\/} satisfying truth assignments, \ie truth
assignments that satisfy, approximately, half of all $km$ literal
occurrences. As it turns out, choosing a concrete range to
represent ``approximately half" and only counting the satisfying
assignments that fall within the range leads to analytic
difficulties due to the polynomial corrections in certain large
deviations estimates. Fortunately, these issues can be avoided by
i) introducing a scheme that weights satisfying truth assignments
according to their number of satisfied literal occurrences, and
ii) tuning the scheme's control parameter so as to concentrate the
weight on balanced assignments.

\subsection{Weighted second moments: a transform}

Recall that for a CNF formula ${\FF}$ on $n$ variables,  ${\cal S}
= {\cal S}({\FF}) \subseteq \{0,1\}^n$ denotes the set of
satisfying truth assignments of ${\FF}$. An attractive feature of
the second moment method is that we are free to apply it to any
random variable $X=X(F)$ such that $X>0$ implies that ${\cal S}
\neq \emptyset$. Sums of the form
$$
X = \sum_{\sigma} w(\sigma,{\FF})
$$
clearly have this property if $w(\sigma,F) = 0$ for $\sigma
\not\in {\cal S}(\FF)$.

{Weighting schemes as above can be viewed as {\em transforms} of
the original problem and can be particularly effective in
exploiting insights into the source of correlations. In
particular, if $w(\sigma,{\FF})$ has product structure over the
clauses, then clause-independence allows one to replace
expectations of products with products of expectations. With this
in mind, let us consider random variables of the form}
\[
X = \sum_{\sigma} \prod_{c} w(\sigma, c) \enspace ,
\]
where $w$ is some arbitrary function. (Eventually, we will require
that $w(\sigma,c)=0$ if $\sigma$ falsifies $c$.) For instance, if
$w(\sigma,c)$ is the indicator that $c$ is satisfied by $\sigma$,
then $X$ simply counts the number of satisfying truth assignments.
By linearity of expectation and clause-independence we see that
for any function $w$,
\begin{eqnarray}
\ex[X] & = & \sum_{\sigma}\prod_{c} \ex[w(\sigma,c)] \enspace , \label{eq:gen1}\\
\ex[X^2] & = & \sum_{\sigma,\tau} \prod_{c}
\ex[w(\sigma,c)\,w(\tau,c)] \enspace . \label{eq:gen2}
\end{eqnarray}

Since we are interested in random formulas where the literals are
drawn uniformly, we will restrict attention to functions that are
independent of the variable labels. That is, for every truth
assignment $\sigma$ and every clause $c = \ell_1 \vee \cdots \vee
\ell_k$, we {require that} $w(\sigma,c) = w(\vv)$, where $v_i=+1$
if $\ell_i$ is satisfied under $\sigma$ and $-1$ if $\ell_i$ is
falsified under $\sigma$. With that assumption, \eqref{eq:gen1}
and \eqref{eq:gen2} simplify to
\begin{eqnarray}
\ex[X] & = & 2^n \left(\ex[w(\sigma,c)]\right)^m \enspace ,\\
\ex[X^2] & = & \sum_{\sigma,\tau} \left(\ex[w(\sigma,c)\,
w(\sigma,\tau)]\right)^m \enspace . \label{fasolakia}
\end{eqnarray}

Let $A = \{-1,+1\}^k$. Since literals are drawn uniformly and
independently we see that for every $\sigma$,
\[
\ex[w(\sigma, c)]  =  \sum_{\vv \in A} w(\vv) \, 2^{-k} \enspace .
\]
Similarly, for every  pair of truth assignments $\sigma,\tau$ with
overlap $z = \alpha n$,
\begin{eqnarray}
\ex[w(\sigma,c)\, w(\sigma,\tau)] & = & \sum_{\uu,\vv \in A}
w(\uu) w(\vv) \;  2^{-k}\prod_{i=1}^k\left(\a^{\one_{u_i =
v_i}}(1-\a)^{\one_{u_i \neq v_i}}\right) \nonumber \\
& \equiv & \sum_{\uu,\vv \in A} w(\uu) w(\vv) \;
\Phi_{\uu,\vv}(\a) \nonumber\\
& \equiv & f_w(\a)  \label{fasola} \enspace .
\end{eqnarray}
In particular, observe that $\ex[w(\sigma, c)]^2 = f_w(1/2)$, \ie
for every function $w$ the weights assigned to truth assignments
with overlap $n/2$ are independent.

Recalling the approximation $\binom{n}{z} =
(\a^\a(1-\a)^{1-\a})^{-n} \times {\mathrm {poly}}(n)$ we see
that~\eqref{fasolakia},\eqref{fasola} imply
\begin{eqnarray}
\ex[X^2] & = & 2^n \sum_{z=0}^n  \binom{n}{z} \,f_w(z/n)^{m}
            \label{hopa}\\
& \leq & 2^n \left(\max_{0 \leq \a \leq 1}
            \left[\frac{f_w(\a)^r}{\a^\a(1-\a)^{1-\a}}\right]
        \right)^n \times {\mathrm {poly}}(n) \nonumber \\
 & \equiv &  \left(\max_{0 \leq \a \leq 1} \Lambda_w(\a)\right)^n
\times {\mathrm {poly}}(n) \enspace . \label{bopa}
\end{eqnarray}
Observe that $\Lambda_w(1/2)^n =
\left(4f_w(1/2)^r\right)^n=\ex[X]^2$. Moreover, we will see later
that a more careful analysis of the sum in~\eqref{hopa} allows one
to replace the polynomial factor in~\eqref{bopa} by $O(1)$.
Therefore, if $\Lambda_w(1/2)$ is the global maximum of
$\Lambda_w$ then $\ex[X^2]/\ex[X]^2 = O(1)$ and the second moment
method succeeds.

A necessary condition for $\Lambda_w(1/2)$ to be a global maximum
is that $\Lambda'_w(1/2)=0$. Since $\Lambda_w(\alpha) =
2\mathcal{E}(\a)f_w(\a)^r$ and $\mathcal{E}'(1/2)=0$, this
dictates $f_w'(1/2)=0$. Differentiating $f_w$ we get
\begin{eqnarray*}
f_w'(\a) & = & \sum_{\uu,\vv \in A} w(\uu) w(\vv) \;
\Phi_{\uu,\vv}(\a) \;
\left[\,\log \Phi_{\uu,\vv}(\a) \, \right]' \\
& = & \sum_{\uu,\vv \in A} w(\uu) w(\vv) \; \Phi_{\uu,\vv}(\a) \;
\sum_{i=1}^k \left(\frac{\one_{u_i = v_i}}{\a}-\frac{\one_{u_i
\neq v_i}}{1-\a}\right) \enspace .
\end{eqnarray*}
In particular, letting $\uu \cdot \vv$ denote the inner product of
$\uu$ and $\vv$, we see that
\begin{equation}\label{eq:balance}
f_w'(1/2) = 2^{-2k+1}\sum_{\uu,\vv \in A} w(\uu) w(\vv) \; \uu
\cdot \vv = \dimf{2^{-2k+1}} \left(\sum_{\uu \in A} w(\uu) \uu
\right) \cdot \left(\sum_{\vv\in A} w(\vv) \vv\right) \enspace .
\end{equation}
Therefore,  for any function $w$
\begin{equation}\label{eq:cool}
f_w'(1/2)=0 \Longleftrightarrow \sum_{\vv \in A} w(\vv) \vv =0
\enspace .
\end{equation}

{We can interpret the vanilla application of the first moment
method as using a function $w=w_S$ which assigns \dimt{0 to
$(-1,\ldots,-1)$ and $1/(2^k-1)$ to all other vectors. (It is
convenient to always normalize $w$ so that $\sum_{\vv} w(\vv)=1$.)
The fact that $w_S$ violates the r.h.s.\ of~\eqref{eq:cool},
implies that this attempt must fail. In~\cite{naesat}, on the
other hand, $w=w_N$ assigns 0 both to $(-1,\ldots,-1)$ and to
$(+1,\ldots,+1)$ (and $1/(2^k-2)$ to all other vectors), thus
satisfying~\eqref{eq:cool} and enabling the second moment method.
Nevertheless, this particular rebalancing of the vectors is rather
heavy-handed since it makes it is twice as likely to assign zero
to a random clause.

To achieve better results we would like to choose a function $w$
that is ``as close as possible" to $w_S$ while
satisfying~\eqref{eq:cool}. That is, we would like $w$ to have
minimal relative entropy with respect to $w_S$ subject
to~\eqref{eq:cool} (see Definition 2.15 of~\cite{dez}). Since
$w_S$ is constant over all $\vv \neq (-1,\ldots,-1)$ \dimss{and we
must have $w(-1,\ldots,-1) = w_S(-1,\ldots,-1)=0$,} this means
that $w$ should have maximum entropy over $\vv \neq
(-1,\ldots,-1)$ while satisfying~\eqref{eq:cool}.} So, all in all,
we are seeking a maximum-entropy collection of weights for the
vectors in $A$ such that i) the all -1s vector has weight 0, ii)
the weighted vectors cancel out.

For $\xx \in A$, let $|\xx|$ denote the number of +1s in $\xx$. By
summing the r.h.s.\ of~\eqref{eq:cool} over the coordinates we see
that a necessary condition for the optimality of $w$ is
    \begin{equation}\label{eq:coord}
    \sum_{\vv \neq (-1,\ldots,-1)} w(\vv)(2|\vv|-k)=0 \enspace .
    \end{equation}
Maximizing entropy subject to~\eqref{eq:coord} is a standard
Lagrange multipliers problem. Its unique solution is
\begin{equation}\label{eq:coorda}
w(\vv) = \frac{1}{Z} \, \lambda^{|\vv|}  \enspace ,
\end{equation}
where $Z$ is a normalizing constant and $\lambda$ satisfies
$(1+\lambda)^{k-1}=1/(1-\lambda)$ so that~\eqref{eq:coord} is
satisfied, \ie
\begin{equation}\label{eq:lasting}
\sum_{j=1}^k \binom{k}{j}  \lambda^j (2j-k)  = \dimf{k
\left(1-(1+\lambda)^{k-1}(1-\lambda)\right)}= 0\enspace .
\end{equation}
Note now that for $w$ given by~\eqref{eq:coorda}, symmetry ensures
that all coordinates of $\sum_{\vv} w(\vv)\vv$ are equal. Since,
by~\eqref{eq:lasting}, the sum over these coordinates vanishes, we
see that in fact~\eqref{eq:cool} must hold as well. Therefore, $w$
is indeed the optimal solution for our original problem.

We plot below the functions $f_w$ and $\Lambda_w$ corresponding to
this weighting, for the values of $k,r$ in Figure~1. {(With a
normalization for $\sum_uw(\uu)$ which makes the plot scale
analogous to that in Figure~1 and which will be more convenient
for computing $f_w$ and $\Lambda_w$ in the next section.)}

\vspace*{1cm}

\begin{figure}[h]\label{fig_smart}
\begin{minipage}{3in}
\begin{center}
  \centerline{\psfig{figure=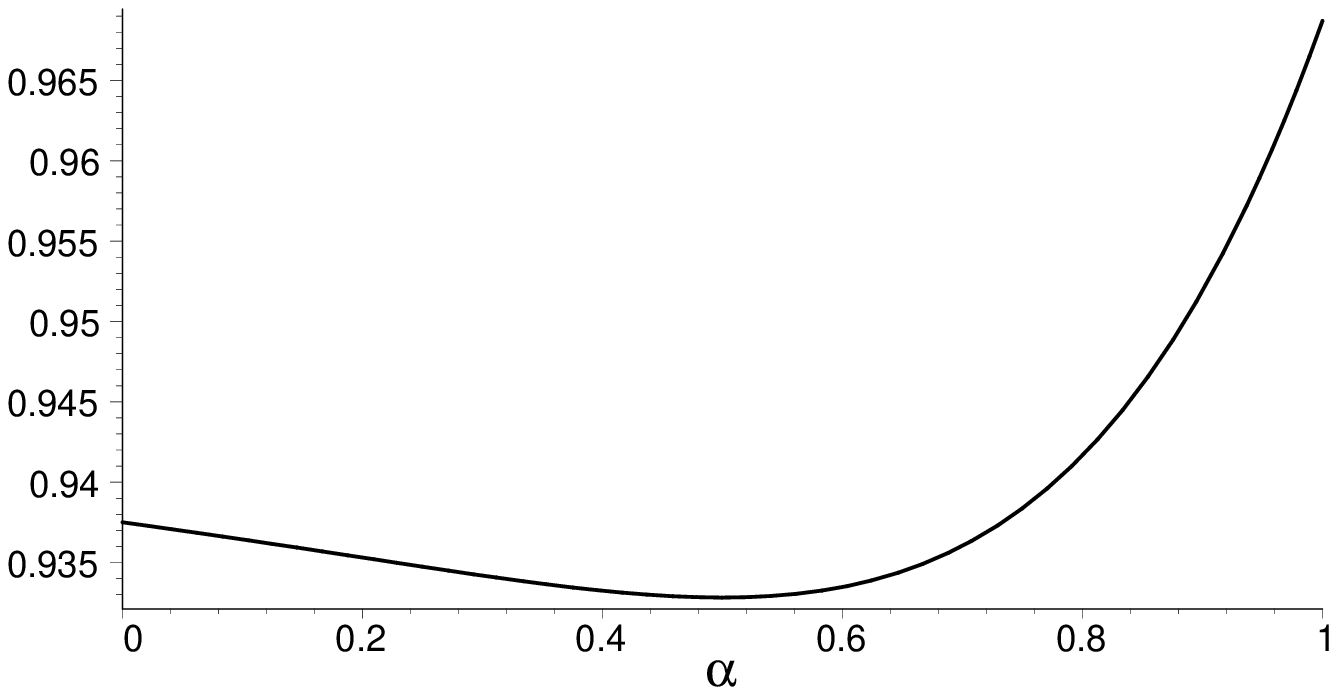,height=1.5in,width=2in}}
  \centerline{$f_w(\alpha)$}
\end{center}
\end{minipage}
\begin{minipage}{3in}
\begin{center}
  \centerline{\psfig{figure=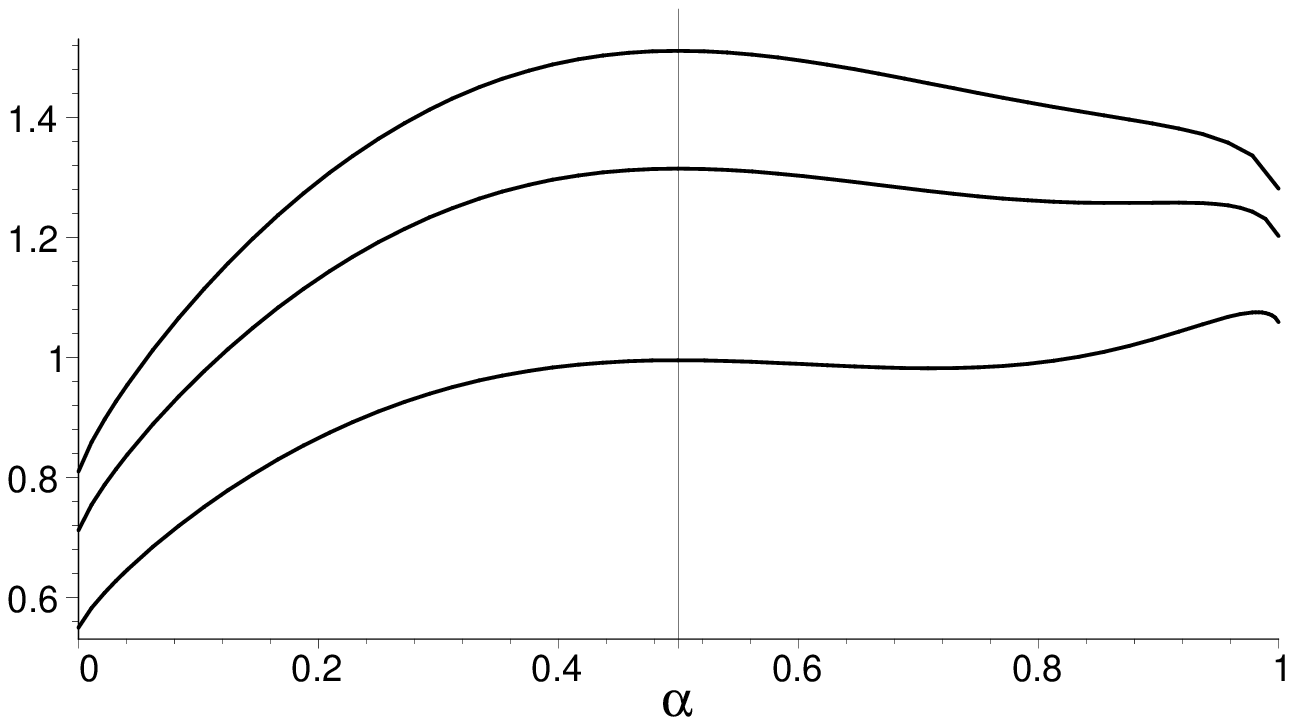,height=1.5in,width=2in}}
  \centerline{$\Lambda_w(\alpha)=2\mathcal{E}(\alpha)f_w(\alpha)^r$}
\end{center}
\end{minipage}
\caption{$k=5$, $r=14,16,20$ (top to bottom).}
\end{figure}

\dims{In conclusion,} if $L(\sigma,F)$ denotes the number of
satisfied literal occurrences in $F$ under $\sigma$, we will take
\begin{equation}\label{eq:propto} w(\sigma,F) \propto
\prod_{c} \lambda^{L(\sigma,c)} \one_{\sigma \in {\cal S}(c)}
\enspace ,
\end{equation}
where $(1+\lambda)^{k-1}=1/(1-\lambda)$. The above weighting
scheme yields Theorem~\ref{easy}, below, which we will prove in
Sections~\ref{ground}---\ref{end_weight}. Theorem~\ref{easy} has
the same leading term as Theorem~\ref{main} but a linear
correction term 4 times greater. This lost factor of 4 is due to
our insistence that $w(\sigma,{\FF})$ factorizes perfectly over
the clauses. In Sections~\ref{trunc}---\ref{end_trunc} we go
beyond what can be achieved with perfect factorization by
performing a truncation. This will allow us to prove
Theorem~\ref{main} which gives a lower bound for $r_k$ that is
within an additive constant of the upper bound for the existence
of balanced satisfying assignments.

\begin{theorem}\label{easy}
There exists a sequence $\beta_k \rightarrow 0$ such that for all
$k \geq 3$,
$$
r_k \geq 2^k \log2 -2(k+1)\log 2 - 1 -\beta_k \enspace .
$$
\end{theorem}

\section{Groundwork}\label{ground}

Given a $k$-CNF formula ${\FF}$ on $n$ variables, recall that
${\cal S}({\FF})$ is the set of satisfying truth assignments of
${\FF}$. Given $\sigma \in \{0,1\}^n$ let $H= H(\sigma,{\FF})$ be
the number of satisfied {literal occurrences} in ${\FF}$ under
$\sigma$ less the number of unsatisfied literal occurrences in
${\FF}$ under $\sigma$. For any $0<\gamma\leq 1$, let
$$
    X =  \sum_{\sigma}   \gamma^{H(\sigma,{\FF})}
    \one_{\sigma \in {\cal S}(\FF)}\enspace .
$$
(Note that $\gamma^{H(\sigma,{\FF})} =
\gamma^{2S(\sigma,{\FF})-km}$, so this is consistent
with~\eqref{eq:propto} for $\gamma^2 = \lambda$.)

Recall that in $F_k(n,m)$ the $m$ clauses $\{c_i\}_{i=1}^m$ are
i.i.d. random variables, $c_i$ being the conjunction of $k$
i.i.d.\ random variables $\{\ell_{ij}\}_{j=1}^k$, each $\ell_{ij}$
being a uniformly random literal. Clearly, in this model a clause
may be improper, \ie it might contain repeated and/or
contradictory literals. At the same time, though, observe that the
probability that a random clause is improper is smaller than
$k^2/n$ and, moreover, the proper clauses are uniformly selected
among all proper clauses. Therefore \whp\ the number of improper
clauses is $o(n)$ implying that if for a given $r$, $F_k(n,rn)$ is
satisfiable \whp\ then for $m=rn-o(n)$, the same is true in the
model where we only select among proper clauses. The issue of
selecting clauses without replacement is completely analogous as
\whp\ there are $o(n)$ clauses that contain the same $k$ variables
as some other clause.

\subsection{The first moment}

For any fixed truth assignment $\sigma$ and a random $k$-clause $c
= \ell_1 \vee \cdots \vee \ell_k$, since the literals
$\ell_1,\ldots,\ell_k$ are i.i.d. we have
\begin{eqnarray*}
    \ex[\gamma^{H(\sigma,c)} \one_{\sigma \in {\cal S}(c)}] & = &
    \ex[ \gamma^{H(\sigma,c)}] - \ex[\gamma^{-k} \one_{\sigma
    \not\in {\cal S}(c)}] \\
& = &     \ex\left[\prod_{\ell_i}\gamma^{H(\sigma,\ell_i)}\right]
-
(2\gamma)^{-k} \\
& = & \left(\frac{\gamma+\gamma^{-1}}{2}\right)^k - (2\gamma)^{-k}
\\
& \equiv & \psi(\gamma) \enspace .
\end{eqnarray*}
Thus, since the $m = rn$ clauses $c_1,c_2,\ldots,c_m$ are i.i.d.\
\begin{eqnarray}
    \ex[X]   & = & \ex \left[\sum_{\sigma}  \gamma^{H(\sigma,{\FF})} \one_{\sigma \in {\cal S}(\FF)}\right] \nonumber \\
            & = & \sum_{\sigma}   \ex\left[\prod_{c_i}
                            \gamma^{H(\sigma,{c_i})} \one_{\sigma \in {\cal S}(c_i)}
                    \right] \nonumber \\
            & = & \sum_{\sigma}   \prod_{c_i}\ex
                    \left[
                            \gamma^{H(\sigma,c_i)} \one_{\sigma \in {\cal S}(c_i)}
                    \right] \nonumber \\
            & = & \left(2 \psi(\gamma)^r\right)^n \enspace . \label{eq:sq_exp_psi}
\end{eqnarray}

\subsection{The second moment}

Let $\sigma,\tau$ be any pair of truth assignments that agree on
$z = \alpha n$ variables. If $\ell_1,\ell_2,\ldots \ell_k$ are
i.i.d.\ uniformly random literals and $c = \ell_1 \vee \ell_2 \vee
\cdots \vee \ell_k$ then
\begin{eqnarray*}
\ex     \left[
            \gamma^{H(\sigma,\ell_i)+H(\tau,\ell_i)}
        \right] & = &  \a \left(\frac{\gamma^2+\gamma^{-2}}{2}\right)+1-\a \enspace ,\\
\ex     \left[
              \gamma^{H(\sigma,\ell_i)+H(\tau,\ell_i)} \one_{\sigma \not\in {\cal S}(c)}
        \right] & = & 2^{-k}\left(\a\gamma^{-2}+(1-\a)\right) \enspace ,\\
\ex     \left[
              \gamma^{H(\sigma,\ell_i)+H(\tau,\ell_i)} \one_{\sigma,\tau \not\in {\cal S}(c)}
        \right] & = & 2^{-k}\left(\a \gamma^{-2}\right) \enspace .
\end{eqnarray*}

Since $\ell_1,\ell_2,\ldots \ell_k$ are i.i.d., writing $\gamma^2
= 1-\e$, we have
\begin{eqnarray}
{\ex \left[
              \gamma^{H(\sigma,c)+H(\tau,c)} \one_{\sigma,\tau \in {\cal S}(c)}
        \right]}
& = &\ex     \left[
              \gamma^{H(\sigma,c)+H(\tau,c)} \left(
                                                    \one
                                                    - \one_{\sigma \not\in  {\cal S}(c)}
                                                    - \one_{\tau   \not\in  {\cal S}(c)}
                                                    + \one_{\sigma,\tau \not\in {\cal S}(c)}
              \right)
        \right] \nonumber \\
& = &\ex   \left[
              \prod_i \gamma^{H(\sigma,\ell_i)+H(\tau,\ell_i)} \!\left(
                                                    \one
                                                    - \one_{\sigma \not\in  {\cal S}(c)}
                                                    - \one_{\tau   \not\in  {\cal S}(c)}
                                                    + \one_{\sigma,\tau \not\in {\cal S}(c)}
              \right)\!
        \right] \nonumber \\
& = & \left(\a \left(\frac{\gamma^2+\gamma^{-2}}{2}\right)+1-\a
    \right)^k - 2^{1-k} \left(\a\gamma^{-2}+(1-\a)\right)^k +
2^{-k}
\left(\a\gamma^{-2}\right)^k \nonumber \\
& = & \frac{(2-2\e+\a\e^2)^k - 2 (1-\e+\a\e)^k+\a^k}{2^k(1-\e)^k} \label{eq:corr}\\
& \equiv  &\frac{f(\alpha)}{2^k(1-\e)^k} \enspace ,
\label{eq:mity}
\end{eqnarray}
where the dependence of $f$ on $\e = 1- \gamma^2$ is implicit.
\dimf{(Taking $\e=1-\lambda$ in~\eqref{eq:corr} yields the $f_w$
of Figure~2.)}

Thus, for a random $k$-CNF formula whose $m = rn$ clauses
$c_1,c_2,\ldots,c_m$ are constructed  independently
\begin{eqnarray}
\ex [X^2]     & = & \ex \left[\sum_{\sigma}  \gamma^{H(\sigma,{\FF})} \one_{\sigma \in {\cal S}(\FF)} \right]^2 \nonumber \\
            & = & \sum_{\sigma,\tau} \ex \left[
              \gamma^{H(\sigma,{\FF})+H(\tau,{\FF})} \one_{\sigma,\tau \in {\cal S}({\FF})}
        \right] \nonumber \\
        & = & \sum_{\sigma,\tau}   \ex\left[\prod_{c_i}
              \gamma^{H(\sigma,{c_i})+H(\tau,{c_i})} \one_{\sigma,\tau \in {\cal S}({c_i})}
                    \right] \nonumber \\
            & = & \sum_{\sigma,\tau}   \prod_{c_i}\ex
                    \left[
              \gamma^{H(\sigma,{c_i})+H(\tau,{c_i})} \one_{\sigma,\tau \in {\cal S}({c_i})}
                    \right] \label{eq:monday}
            \enspace.
\end{eqnarray}

Since the number of ordered pairs of assignments with overlap $z$
is $ 2^n \,\binom{n}{z} $ and since the $m=rn$ clauses are
identically distributed, \eqref{eq:monday} and~\eqref{eq:mity}
imply
\begin{equation}
 \ex[X^2] = 2^n \sum_{z=0}^n  \binom{n}{z} \left(\frac{f(z/n)}{2^k(1-\e)^k}\right)^{rn}
 \enspace .
\label{eq:naesum}
\end{equation}

Observe now that for, any fixed value of $\e$,  $f^r$ is a real,
positive and twice-differentiable. Thus, to bound the sum
in~\eqref{eq:naesum} we can use the following lemma
of~\cite{naesat}. The idea is that sums of these type are
dominated by the contribution of $\Theta(n^{1/2})$ terms around
the maximum term and the proof follows by applying the Laplace
method of asymptotic analysis~\cite{debruijn}.
\begin{lemma}\label{lem:peak}
Let $\phi$ be any real, positive, twice-differentiable function on
$[0,1]$ and let
$$
S_n = \sum_{z=0}^n \binom{n}{z} \,\phi(z/n)^n \enspace .
$$
Letting $0^0 \equiv 1$, define $g$ on $[0,1]$ as
\[
g(\alpha) = \frac{\phi(\alpha)}
                 {\alpha^\alpha \,(1-\alpha)^{1-\alpha}} \enspace
                 .
\]
If there exists $\amax \in (0,1)$ such that $g(\amax) \equiv \gmax
> g(\alpha)$ for all $\alpha \neq \amax$, and $g''(\amax) < 0$,
then there exist constants $B, C > 0$ such that for all
sufficiently large $n$
\[
B \times \gmax^n \,\le\, S_n \,\le\, C \times \gmax^n \enspace .
\]
\end{lemma}

With Lemma~\ref{lem:peak} in mind, let us define
\begin{equation}\label{eq:gr_def}
g_r(\alpha) =
\frac{f(\alpha)^r}{\alpha^{\alpha}(1-\alpha)^{1-\alpha}} \enspace
.
\end{equation}
Let
$$
s_k = 2^k \log2 -2\log 2(k+1) - 1 -3/k \enspace .
$$
We will prove that
\begin{lemma}\label{gleas}
Let $\e$ be such that
\begin{equation}\label{eq:studly}
\e(2-\e)^{k-1}=1 \enspace .
\end{equation}
For all $k \geq 22$, if $r \leq s_k$ then $g_r(1/2)>g_r(\alpha)$
for all $\alpha \neq 1/2$, and $g_r''(1/2) < 0$.
\end{lemma}

As a result, for $r,k,\e$ as in Lemma~\ref{gleas} we have
\begin{equation}\label{eq:taco}
\ex[X^2] < C \times \left(\frac{2
g_r(1/2)}{(2(1-\e))^{kr}}\right)^n \enspace ,
\end{equation}
where $C=C(k)$ is  independent of $n$. Observe now
that~\eqref{eq:sq_exp_psi} and the fact $\gamma^2 = 1-\e$ imply
\begin{eqnarray}
\ex[X]^2 & = & \left[(2\psi(\gamma)^{r})^{n}\right]^2 \nonumber \\
& = & 4^n \left(\frac{f(1/2)}{2^k(1-\e)^k}\right)^{rn} \nonumber \\
& = & \left(\frac{2g_r(1/2)^n}{(2(1-\e))^{kr}}\right)^n
\label{eq:sq_exp} \enspace.
\end{eqnarray}
Therefore, by~\eqref{eq:taco} and~\eqref{eq:sq_exp} we see that
for $r,k,\e$ as in Lemma~\ref{gleas} we have
$$
  {\ex[X^2]} < C \times  {\ex[X]^2}  .
$$
By Lemma~\ref{lemma:sec}, this implies $ \Pr[X > 0] > 1/C $ and,
hence, Lemma~\ref{gleas} along with Corollary~\ref{cor:boost}
imply Theorem~\ref{easy}.\bigskip

To prove Lemma~\ref{gleas} we will prove the following three
lemmata. The first lemma holds for any $\e\in [0,1)$ and reduces
the proof to the case $\alpha \geq 1/2$. The second lemma controls
the behavior of $f$ (and thus $g_r$) around $\a = 1/2$ and demands
the judicious choice of $\e$ specified by~\eqref{eq:studly}. We
note that this is the only value of $\e$ for which $g_r$ has a
local maximum at $1/2$, for any $r>0$. The third lemma deals with
$\alpha$ near 1. That case needs to be handled separately because
$g_r$ has another local maximum in that region. The condition $r
\leq s_k$ aims precisely at keeping the value of $g_r$ at this
other local maximum smaller than $g_r(1/2)$.

\begin{lemma}\label{lem:left_half}
For all $\e,x>0$,  $\, g_r(1/2+x) > g_r(1/2-x)$.
\end{lemma}

\begin{lemma}\label{lem:first_part}
Let $\e$ satisfy \eqref{eq:studly}. For all $k \geq 22$, if $r
\leq 2^k \log 2$ then $g_r(1/2) > g_r(\a)$ for all $\a \in
(1/2,4/5]$ and $g_r''(1/2)<0$.
\end{lemma}

\begin{lemma}\label{lem:sec_part}
Let $\e$ satisfy \eqref{eq:studly}. For all $k \geq 22$, if $r
\leq s_k$ then $g_r(1/2) > g_r(\a)$ for all $\a \in (4/5,1]$.
\end{lemma}

The following bound will be useful. If $\e$ satisfies
\eqref{eq:studly}, then
\begin{equation}\label{eq:eps_bound}
2^{1-k} + k4^{-k} < \e <  2^{1-k} + 3k4^{-k} \enspace .
\end{equation}
To prove~\eqref{eq:eps_bound} let $q(x) = x - 1/(2-x)^{k-1}$ and
observe that  for all $k \geq 3$, the quantity
$q(2^{1-k}+ck4^{-k})$ is negative for $c=1$ but positive for
$c=3$.

\section{Proof of Lemma~\ref{lem:left_half}}

Observe that $\a^{\a}(1-\a)^{1-\a}$ is symmetric around $1/2$ and
that $r>0$. Therefore, it suffices to prove that $f(1/2+x) >
f(1/2-x)$, for all $x > 0$. To do this we first note that
\dima{for all $x \neq 0$,}
\begin{eqnarray}
{2^k f(1/2+x)}
& = & \left((2-\e)^2+2x\e^2\right)^k-2\left(2-\e+2x\e\right)^k+(1+2x)^k\nonumber\\
& = & \sum_{j=0}^k \binom{k}{j} \left[(2-\e)^{2(k-j)}(2x\e^2)^{j}-2(2-\e)^{k-j}(2x\e)^{j}+(2x)^{j}\right]\nonumber\\
& = &
\sum_{j=0}^k \binom{k}{j}(2x)^{j}\left[(2-\e)^{2(k-j)}\e^{2j}-2(2-\e)^{k-j}\e^{j}+1\right]\nonumber\\
& = & \sum_{j=0}^k\binom{k}{j}(2x)^{j}[(2-\e)^{k-j}\e^{j}-1]^2
\enspace. \label{eq:f_pos}
\end{eqnarray}
Thus,  for all $x> 0$,
\begin{equation*}
f(1/2+x) - f(1/2-x) \; = \;
2^{-k}\sum_{j=0}^k\binom{k}{j}2^j[(2-\e)^{k-j}\e^j-1]^2
(x^j-(-x)^j) \; > \; 0 \enspace.
\end{equation*}

\section{Proof of Lemma~\ref{lem:first_part}}

We will prove that for all $k \geq 22$ and $r \leq 2^k \log 2$,
$g_r$ is strictly decreasing in $(1/2,4/5]$. We have
\begin{flalign}
f'(\alpha)  &=   k
\left[(2-2\e+\a\e^2)^{k-1}\e^2-2(1-\e+\a\e)^{k-1}\e+\a^{k-1}\right], \nonumber \\
\nonumber \\
g'_r(\alpha) & =  \frac{f(\a)^{r-1}  \left(r f'(\alpha)+
f(\alpha)(\log(1-\a)-\log\a)\right)}{\a^{\a}(1-\a)^{1-\a}}
\enspace . \label{eq:gdert}
\end{flalign}
So, $ f'(1/2) = k2^{-k+1}\left((2-\e)^{k-1}\e-1\right)^2 $ and
since, by \eqref{eq:studly}, we have $(2-\e)^{k-1}\e=1$ we get
\begin{equation}\label{eq:reme}
g'_r(1/2) \; = \;  f'(1/2) \; = \; 0 \enspace .
\end{equation}

Since $g'_r(1/2)=0$ and, by~\eqref{eq:f_pos}, $f(\a)>0$ for all
$\a$ we see that \eqref{eq:gdert} implies that to prove that $g_r$
is decreasing in $(1/2,4/5]$ it suffices to prove that the
derivative of
\begin{equation}\label{eq:bart}
r f'(\alpha) +f(\alpha)(\log(1-\a)-\log\a)
\end{equation}
is negative in $(1/2,4/5]$. We will actually prove this claim for
$\alpha \in [1/2,4/5]$. Since $f'(1/2)=0$ this  also establishes
our claim that $g''_r(1/2)<0$. The derivative of~\eqref{eq:bart}
is
\begin{equation}\label{eq:sec_der}
 rf''(\alpha)+ f'(\alpha)(\log(1-\a)-\log\a)-f(\a)\left(\frac{1}{\a}+\frac{1}{1-\a}\right)
\enspace .
\end{equation}
By considering~\eqref{eq:f_pos}, we see that $f$ is non-decreasing
in $[1/2,1]$. Since $\log(1-\a)\leq\log\a$ for $\alpha \in
[1/2,1)$, it follows that in order to prove that the expression
in~\eqref{eq:sec_der} is negative it suffices to show that
$$
r f''(\alpha)\leq f(\a)\left(\frac{1}{\a}+\frac{1}{1-\a}\right)
\enspace .
$$

Since, by definition, $\e < 1$ it follows that $\a\e^2\leq 2\e$
implying that we can bound $f''$ as
\begin{eqnarray}
f''(\alpha)  &=&
{k(k-1)\left((2-2\e+\a\e^2)^{k-2}\e^4-2(1-\e+\a\e)^{k-2}\e^2+\a^{k-2}\right)}
\nonumber\\
& \leq & k^2
\left(2^{k-2}\e^4+(4/5)^{k-2}\right)\label{eq:uselater} \enspace .
\end{eqnarray}
At the same time, $1/\a+1/(1-\a) \geq 4$ and $f(\alpha) \geq
f(1/2)= 2^{-k}((2-\e)^k-1)^2$. Therefore, if $\e_u$ is any upper
bound on $\e$ it suffices to establish
\begin{equation}\label{eq:r_cond1}
r  \times k^2 \left(2^{k-2}\e_u^4+(4/5)^{k-2}\right) \leq 4 \times
2^{-k}((2-\e_u)^k-1)^2 \enspace .
\end{equation}
Invoking~\eqref{eq:eps_bound} to take $\e_u = 2^{1-k} + 3k4^{-k}$,
it is easy to verify that~\eqref{eq:r_cond1} holds for $k \geq 22$
and $r = 2^k \log 2$.\hfill$\Box$\medskip

\begin{corollary}\label{cor:nine}
  For all $k \geq
65$, if $r \leq 2^k \log 2$ then $g_r(1/2) > g_r(\a)$ for all $\a
\in (1/2,9/10]$ and $g_r''(1/2)<0$.
\end{corollary}
\noindent{\bf Proof.} If in~\eqref{eq:r_cond1} we replace $4/5$
with $9/10$ and take $r=2^k \log 2$, then the inequality is valid
for all $k \geq 65$. \bigskip

\section{Proof of Lemma~6}
\label{end_weight}

First observe that the inequality $g_r(1/2) > g_r(\a)$ is
equivalent to
\begin{equation}\label{neq:est}
        \left(\frac{f(\a)}{f(1/2)}\right)^r < 2
        \a^{\a}(1-\a)^{1-\a} \enspace .
\end{equation}
Recall now that, by~\eqref{eq:f_pos}, $f$ is increasing in
$(1/2,1]$ implying $f(\alpha) - f(1/2)> 0$ and that for all $x\geq
0$, $\log(1+x)\leq x$. Thus, the logarithm of the left hand side
above can be bounded as
\begin{eqnarray*}
r \log\left(\frac{f(\a)}{f(1/2)}\right) & = & r \log \left(1 +
\frac{f(\a)-f(1/2)}{f(1/2)}\right)\\
& \leq & r \left(\frac{f(\a)-f(1/2)}{f(1/2)}\right) \enspace .
\end{eqnarray*}
So, if we let $h(\a) = -\a\log\a - (1-\a)\log(1-\a)$ we see
that~\eqref{neq:est} holds if
$$ r < (\log 2 -
h(\a)) \times \frac{f(1/2)}{f(\a)-f(1/2)} \enspace .
$$

To get a lower bound on $f(1/2)$ we use the upper bound for $\e$
from~\eqref{eq:eps_bound}, yielding
\begin{eqnarray}
f(1/2) & = & (2-2\e+\e^2/2)^k - 2 (1-\e/2)^k+(1/2)^k \nonumber \\
& > & (2(1-\e))^k - 2  \nonumber \\
& > & 2^k(1-k\e) -2  \nonumber\\
& > & 2^k(1-k(2^{1-k} + 3k4^{-k})) -2 \nonumber \\
& = & 2^k - 2k - 2 - 3k^22^{-k} \label{neq:halflife}\enspace .
\end{eqnarray}

To get an upper bound on $f(\alpha)-f(1/2)$ we let $\alpha=1/2+x$
and consider the sum in~\eqref{eq:f_pos}. By our choice of $\e$
in~\eqref{eq:studly} we see that: i) the term corresponding to
$j=1$ vanishes yielding~\eqref{eq:first two}, and ii) for all
$j>1$, $0< (2-\e)^{k-j} \e^j<1$ yielding~\eqref{eq:rest_terms}.
That is,
\begin{eqnarray}
f(1/2+x) & = &  f(1/2) + 2^{-k}\sum_{j=2}^k\binom{k}{j}(2x)^{j}[(2-\e)^{k-j}\e^{j}-1]^2 \label{eq:first two} \\
& \leq  & f(1/2) + 2^{-k}\sum_{j=0}^k\binom{k}{j}(2x)^{j} \label{eq:rest_terms}  \\
& = & f(1/2) + \alpha^k \enspace \label{eq:last}.
\end{eqnarray}

Therefore, we see that~\eqref{neq:est} holds as long as
$$ r \leq \frac{\log 2 -
h(\a)}{\a^k} \times f(1/2) \equiv \phi(\a)\times f(1/2) \enspace .
$$

We start by getting a lower bound for $\phi$ for all $\alpha \in
(1/2,1]$. For that, we let $y=1-\alpha$ and observe that for all
$0 < y \leq 1/2$
\begin{equation}\label{eq:entr}
-h(1-y) \; > \; \log(1-y) + y \log y \; > \;
 - y - y^2 + y \log y
\end{equation} and
$$
\frac{1}{(1-y)^k} > (1+y)^k > 1+ky \enspace .
$$
Therefore,
\begin{eqnarray}
\phi(1-y) & = & \frac{\log 2 - h(1-y)}{(1-y)^k} \nonumber \\
& > & (1+ky) (\log 2 - y(1+y-\log y)) \label{eq:generic} \enspace
.
\end{eqnarray}
Writing $y = d/2^k$ and substituting into~\eqref{eq:generic} we
get that for all $1/2 \leq \a< 1$, \dimf{\begin{eqnarray}
\phi(\alpha) & = & \phi(1-d2^{-k}) \nonumber \\
& > &
 \left(1+kd2^{-k}\right) \left(\log 2 - d2^{-k}\left(1+d2^{-k}-\log(d2^{-k})\right)\right) \nonumber\\
& =  &\log 2 + {d(\log d -1)}2^{-k}-\frac{d^2}{4^{k}}\left(1+k\left(1+d2^{-k}-\log(d2^{-k})\right)\right) \nonumber\\
& \geq  &\log 2 - 2^{-k}-\frac{d^2}{4^{k}}\left(1+k\left(1+d2^{-k}-\log(d2^{-k})\right)\right) \nonumber\\
& =  & \log 2 - 2^{-k} - (1-\a)^2
\left(
    1+k\left(
            2-\a-\log(1-\a)
        \right)
\right) \nonumber\\
& \equiv & b(\a)\label{eq:biggie}
 \enspace .
\end{eqnarray}}

Since $\phi$ is {differentiable}, to bound it in $(4/5,1]$ it
suffices to consider its value at $4/5,1$ and wherever
\begin{equation} \label{eq:phiprime}
\phi'(\a)  = \frac{\a \log\a - \a\log(1-\a) - k \log 2 + k
h(\a)}{\a^{k+1}} = 0 \enspace .
\end{equation}
We start by observing that for $k \ge 6$
\[ \phi'(4/5) < 0 \enspace .\]
At the other end, we see that
\[ \lim_{\alpha \to 1}
\frac{\phi'(\a)}{\log(1-\a)} = -1 \enspace ,\] implying that the
derivative of $\phi$ becomes positively infinite as we approach 1.
Therefore, we can limit our search to the interior of $(4/5,1)$
for $k\geq 6$.

By setting $\phi'$ to zero, \eqref{eq:phiprime} gives
$$
\log(1-\a) = \log\a - \frac{k \log 2 - k h(\a)}{\a}
$$ which, since $1/2<\a<1$, implies
\begin{equation}\label{eq:upperln}
\log(1-\a) \leq  - k (\log 2 - h(\a)) \enspace .
\end{equation}
Moreover, \dimf{since $\log 2 - h(4/5)>1/6$,} we see that
\eqref{eq:upperln} implies $\a
> 1-e^{-k/6}$ for all $k$.
Note now that if $\a > 1-e^{-ck}$ for any $c>0$
then~\eqref{eq:entr} implies $h(\a) < e^{-ck}(1+e^{-ck}+ck)$.
Since $\a > 1- e^{-k/6}$, we thus get
\begin{eqnarray}
h(\a) \; < \; e^{-k/6}(1+e^{-k/6}+k/6)  \; < \;e^{-k/6}(2+k/6) \;
\equiv \;   Q(k) \enspace . \label{eq:entr_a}
\end{eqnarray}
Plugging~\eqref{eq:entr_a} into~\eqref{eq:upperln}, we conclude
that
\begin{equation}\label{eq:r_last}
\a > 1 -  e^{-k(\log 2 - Q(k))} \equiv \a^*_k\enspace .
\end{equation}
\dimf{Since for $k \geq 12$ we have $\a^*_k>4/5$, this means that
$\phi$ is decreasing in $(4/5,\a^*_k]$ for $k \geq 12$.

Note now that the function $b$ bounding $\phi$ from below
in~\eqref{eq:biggie} is increasing in $[0,1]$. Combined with the
fact that $\phi$ is decreasing in $(4/5,\a^*_k]$ this implies that
$b(\a^*_k)$ is a lower bound for $\phi$ in $(4/5,1]$, \ie
\begin{eqnarray}
\phi(\a) & > & b(\a^*_k) \nonumber \\
     & > & \log 2 - 2^{-k} - 2/(k2^{k}) \label{eq:ln2b}\enspace ,
\end{eqnarray}
where~\eqref{eq:ln2b} holds for all $k\geq 22$.}
Combining~\eqref{eq:ln2b} with~\eqref{neq:halflife}, we get that
for all $k\geq 22$ if
$$
r < 2^k \log2 -2\log2(k+1) - 1 -3/k
$$
then $g(1/2)>g(\a)$ for all $\a \neq 1/2$.

\section{Further refinement: truncation and weighting}\label{trunc}
Given a $k$-CNF formula ${\FF}$ on $n$ variables, recall that
${\cal S} = {\cal S}({\FF}) \subseteq \{0,1\}^n$ is the set of
satisfying truth assignments of ${\FF}$. Recall also that for
$\sigma \in \{0,1\}^n$, by $H(\sigma,{\FF})$ we denote the number
of satisfied {\em literal occurrences} in ${\FF}$ under $\sigma$
less the number of unsatisfied literal occurrences. Let ${\cal
S^+}= \{ \sigma \in {\cal S} : H(\sigma, {\FF}) \geq 0\}$.

For any $0<\gamma\leq 1$ let
\begin{eqnarray*}
    X           & = & \sum_{\sigma \in {\cal S}}   \gamma^{H(\sigma,{\FF})} \enspace, \\
    X_+         & = & \sum_{\sigma \in {\cal S^+}} \gamma^{H(\sigma,{\FF})} \enspace .
\end{eqnarray*}

In computing the second moment of $X$ {in the previous sections},
{it becomes clear that one needs to control the contribution to
$\ex[X^2]$ from pairs of truth assignments with high overlap.
Close examination} of these {pairs} shows that the {dominant}
contributions come from {those pairs amongst them} that have {\em
fewer\/} than half of their literals satisfied. If we compute the
second moment of $X_+$ instead, these highly correlated pairs are
avoided. Our argument {for this} is motivated by Cramer's
classical ``change of measure'' technique in large deviation
{theory.}\medskip

 Specifically, let $\e_0<1$ satisfy
\begin{equation}\label{eq:newstudly}
\e_0 = \frac{1}{(2-\e_0)^{k-1}} \enspace .
\end{equation}
Lemma~\ref{doobya} below asserts that \dima{if $\gamma^2 = 1 -
\e_0$, where $\e_0$ is specified by~\eqref{eq:newstudly}, then the
first moments of $X$ and $X_+$ are comparable.}
\begin{lemma}\label{doobya}
If $\gamma^2 = 1 - \e_0$ then as $n \rightarrow \infty$
$$
\frac{\ex [X_+]}{\ex [X]} \rightarrow 1/2 \enspace .
$$
\end{lemma}

Let $\sigma,\tau$ be any pair of truth assignments that agree on
$z = \alpha n$ variables. If we write $\theta^2 = 1-\e$ then
from~(\ref{eq:corr}) we have
\begin{eqnarray}
{\ex\left[\theta^{H(\sigma,c)+H(\tau,c)} \one_{\sigma,\tau
        \in {\cal S}(c)}\right]}
& = & \frac{(2-2\e+\a\e^2)^k - 2
        (1-\e+\a\e)^k+\a^k}{2^k(1-\e)^k} \nonumber \\
& \equiv & f(\a,\e) \enspace . \label{newf0}
\end{eqnarray}
{(Observe that the function $f(\a,\e)$ in~\eqref{newf0} above is
identical to $f(\a)(2(1-\e))^{-k}$, where $f(\a)$ is as
in~\eqref{eq:mity}. In the earlier sections, since $\e$ was fixed,
this dependence on $\e$ was suppressed to simplify notation.)}

Thus, if ${\FF}$ is a random formula consisting of $m=rn$
independent clauses then for any $\theta^2 = 1 - \e \geq
\gamma^2$,
\begin{eqnarray}
{\ex\left[\gamma^{H(\sigma,{\FF})+H(\tau,{\FF})}
   \one_{\sigma,\tau \in {\cal S^+}({\FF})}\right]}
   &\le&
   \ex\left[\theta^{H(\sigma,{\FF})+H(\tau,{\FF})}
   \one_{\sigma,\tau \in {\cal S^+}({\FF})}\right] \nonumber
   \\
   &\le&
    \ex\left[\theta^{H(\sigma,{\FF})+H(\tau,{\FF})}
   \one_{\sigma,\tau \in {\cal S}({\FF})}\right] \nonumber \\
    &=& f(\a,\e)^m   \enspace .\label{newf}
\end{eqnarray}

The crucial point is that~\eqref{newf} holds for any $\e \leq
1-\gamma^2$, allowing us to optimize $\e$ with respect to $\a$. In
particular if $\gamma^2 = 1 - \e_0$ then~\eqref{newf} implies
\begin{eqnarray*}
\ex\left[\gamma^{H(\sigma,{\FF})+H(\tau,{\FF})} \one_{\sigma,\tau
\in {\cal S^+}({\FF})}\right] & \le & \left[\inf_{\e \le \e_0}
   f(z/n,\e)\right]^m
   \enspace .
\end{eqnarray*}
Thus, following the derivation of~\eqref{eq:naesum}, we deduce
that
\begin{equation}
 \ex[X_+^2] \le  2^n \sum_{z=0}^n  \binom{n}{z} \left[\inf_{\e \le \e_0}
   f(z/n,\e)\right]^{rn}
 \enspace .
\label{eq:naesum2}
\end{equation}

{Let us define}
$$
g_r(\a,\e) = \frac{f^r(\a,\e)}{\a^{\a}(1-\a)^{1-\a}} \enspace.
$$
Observe that by Lemma~\ref{doobya} and~\eqref{eq:sq_exp},
\begin{equation}
3 \ex[X_+]^2 > \ex[X]^2 = g_r(1/2,\e_0)^n \enspace.
\end{equation}

\dimf{Assume now that there exists a piecewise-constant function
$\xi$ such that for some value of $r$ we have $g_r(1/2,\e_0) >
g_r(\alpha,\xi(\alpha))$ for all $\alpha \neq 1/2$. Then, by
decomposing the sum in~\eqref{eq:naesum2} along the pieces of
$\xi$ and applying Lemma~\ref{lem:peak} to each piece, we can
conclude that $\ex[X_+^2] < C \times \ex[X_+]^2$, for some $C =
C(k)$.} Lemma~\ref{lemma:sec} and Corollary~\ref{cor:boost} then
imply $r_k \geq r$.

Let
$$
\rho_k = 2^k \log 2 - \frac{\log 2}{2} (k+1) - 1 - 50 k^3 \,2^{-k}
\enspace .
$$
We will prove
\begin{lemma}\label{hardgleas} Let
$$
\xi(\alpha) =
\begin{cases}
\e_0 & \mbox{ if $\alpha \in [1/10,9/10]$}\\
\\
\e_0/2 & \mbox{ otherwise.}
\end{cases}
$$
For all $k \geq 166$, if $r \leq \rho_k$ then
$g_r(1/2,\e_0)>g_r(\alpha,\xi(\alpha))$ for all $\alpha \neq 1/2$,
and the second derivative of $g_r$ with respect to $\alpha$ is
negative at $\alpha=1/2$.
\end{lemma}

To prove Lemma~\ref{hardgleas} we first observe that since $\xi$
is symmetric around $1/2$, Lemma~\ref{lem:left_half} implies that
we only need to consider the case $\alpha \geq 1/2$. Also, since
$\xi(\a) = \e_0$ for $\alpha \in [1/2,9/10]$,
Corollary~\ref{cor:nine} establishes both our claim regarding the
second derivative of $g_r$ at $\alpha=1/2$ and
$g_r(1/2,\e_0)>g_r(\alpha,\xi(\alpha))$ for  $\alpha \in
(1/2,9/10]$.  Thus, besides Lemma~\ref{doobya},  it suffices to
prove that
\begin{lemma}\label{lem:hard_sec_part} For all $k \geq 166$, if $r \leq \rho_k$
then for all $\a \in (9/10,1]$ $ g_r(1/2,\e_0) > g_r(\a,\e_0/2)$.
\end{lemma}

\section{Proof of Lemma~\ref{doobya}}\label{sec:doobya}

By linearity of expectation, it suffices to prove that for
$\gamma^2 = 1-\e_0$ and every $\sigma$
\begin{equation} \label{favor}
    \frac{\ex[\gamma^{H(\sigma,{\FF})}
\one_{\sigma \in {\cal S^+({\FF})}}]}
         {\ex[\gamma^{H(\sigma,{\FF})} \one_{\sigma \in {\cal S}({\cal
         F})}]} \rightarrow \frac{1}{2} \enspace .
\end{equation}

Recalling that formulas in our model are sequences of i.i.d.\
random literals $\ell_1,\ldots,\ell_{km}$, let ${\mathbf
P}(\cdot)$ denote the probability assigned by our distribution to
any such sequence, \ie $(2n)^{-km}$. Now, fix any truth assignment
$\sigma$ and consider an auxiliary distribution ${\mathbf
P}_{\gamma}$ on $k$-CNF formulas where the $km$ literals are again
i.i.d., but where now for each fixed literal $\ell$
$$
{\mathbf P}_{\gamma}[H(\sigma,\ell) = 1] =
\frac{\gamma}{\gamma+\gamma^{-1}} \enspace .
$$

Observe that since $\gamma \leq 1$ this probability is at most
$1/2$. Thus,
$$
\ex_{\gamma} [H(\sigma,\ell)]\; = \;
\frac{\gamma-\gamma^{-1}}{\gamma+\gamma^{-1}} \; = \;
\frac{\gamma^2-1}{\gamma^2+1}\; = \; \frac{-\e_0}{2-\e_0}\enspace
.
$$
So, for a random $k$-clause $c$
\begin{eqnarray*}
\ex_{\gamma}[ H(\sigma,c) \one_{\sigma \in {\cal S}(c)}]  & = &
\ex_{\gamma} [H(\sigma,c)] - \ex_{\gamma}[-k \one_{\sigma \not\in
{\cal S}(c)}] \\
& = & \frac{-k \e_0}{2-\e_0} + k
\left(\frac{\gamma^{-1}}{\gamma+\gamma^{-1}}\right)^k \\
& = & k \left(- \frac{\e_0}{2-\e_0} +
\left(\frac{1}{2-\e_0}\right)^k\right) \enspace .
\end{eqnarray*}
Since $\e_0 = 1/(2-\e_0)^{k-1}$ we see that $\ex_{\gamma}[
H(\sigma,c) \one_{\sigma \in {\cal S}(c)}]=0$.

By literal independence, for any specific clause $c_0$
\begin{equation} \label{doob}
{\mathbf P}_{\gamma}(c_0)= \frac{\gamma^{H(\sigma,c_0)}{\mathbf
P}(c_0)} {(\gamma+\gamma^{-1})^k} \,.
\end{equation}
Let $Z(\gamma)=\ex_{\gamma}[ \one_{\sigma \in {\cal S}(c)}]$ and
$Z_1(\gamma)=Z(\gamma)(\gamma+\gamma^{-1})^k$. For any clause
$c_0$, define
\begin{equation} \label{doob2}
\widetilde{\mathbf P}_{\gamma}(c_0) =
\frac{{\mathbf P}_{\gamma}(c_0) \one_{\sigma \in {\cal S}(c_0)}
}{Z(\gamma)} = \frac{\gamma^{H(\sigma,c_0)}{\mathbf P}(c_0)
\one_{\sigma \in {\cal S}(c_0)}}{Z_1(\gamma)} \enspace ,
\end{equation}
where the second equality follows from (\ref{doob}).
Now pick $m$ i.i.d.\ clauses with \dima{the distribution
in~\eqref{doob2}. Any  fixed formula ${\FF}_0$} will be obtained
with probability
\begin{equation} \label{doob3}
\widetilde{\mathbf P}_{\gamma}({\FF}_0) =
\frac{\gamma^{H(\sigma,{\FF}_0)}{\mathbf P}({\FF}_0) \one_{\sigma
\in {\cal S}({\FF}_0)}}{Z_1(\gamma)^m} \enspace .
\end{equation}

Since  $\widetilde{\mathbf E}_{\gamma}[H(\sigma,c)] = 0$, the
central limit theorem yields,
$$
\widetilde{\mathbf P}_{\gamma}[H(\sigma,{\FF}) \geq 0]
 \rightarrow \frac{1}{2}
$$
as  $n \rightarrow \infty$. By (\ref{doob3}), this is equivalent
to (\ref{favor}).

\section{Proof of Lemma~\ref{lem:hard_sec_part}}\label{end_trunc}

Write $\e_1 = \e_0/2$ (to simplify notation). Observe that the
inequality $g_r(1/2,\e_0) > g_r(\a,\e_1)$ is equivalent to
\begin{equation}\label{neq:hard_est}
        \left(\frac{f(\a,\e_1)}{f(1/2,\e_0)}\right)^r < 2
        \a^{\a}(1-\a)^{1-\a} \enspace .
\end{equation}
If $h(\a) = -\a\log\a - (1-\a)\log(1-\a)$ denotes the entropy
function, then~\eqref{neq:hard_est} is equivalent to
$$ r < \frac{\log 2 -
h(\a)}{\log(1+w)} \enspace ,
$$
where
$$
w = \frac{f(\a,\e_1) - f(1/2,\e_0)}{f(1/2,\e_0)} \enspace .
$$
We will prove $f(\a,\e_1) - f(1/2,\e_0) < 2^{1-k}$. Therefore,
recalling that \dima{for all $x>-1$,}
$$
\frac{1}{\log(1+x)} \geq \frac{1}{x}+\frac{1}{2}-\frac{x}{12}
\enspace ,
$$
we see that~\eqref{neq:hard_est} holds if
\begin{equation}\label{eq:taste}
\frac{r}{\log 2 -h(\a)} <\frac{f(1/2,\e_0)}{f(\a,\e_1) -
f(1/2,\e_0)} +\frac{1}{2} -\frac{1}{2^k f(1/2,\e_0)} \enspace .
\end{equation}\medskip

To get a lower bound on $f(1/2,\e_0)$ we use the upper bound for
$\e_0$ from~\eqref{eq:eps_bound}. Thus, for all $k\geq 5$
\begin{align}
2^k f(1/2,\e_0) & =  \frac{(2-2\e_0+\e_0^2/2)^k - 2 (1-\e_0/2)^k+(1/2)^k}{(1-\e_0)^k} \nonumber \\
& >  \frac{(2-2\e_0)^k - 2}{(1-\e_0)^k}   \nonumber \\
& =  2^k - \frac{2}{(1-\e_0)^k}  \nonumber\\
& >  2^k - 2 - \dima{2(1+k\e_0)}\nonumber \\
& >  2^k - 2 - \dima{k2^{-k+1}} \label{neq:hard_halflife}\enspace
.
\end{align}

To get an upper bound on the numerator of $w$ we let
$\alpha=1/2+x$ and consider the sum in~\eqref{eq:f_pos}
\dima{(recall that~\eqref{eq:f_pos} holds for all $\e$ and that
$f(\a)$ in~\eqref{eq:f_pos} is merely $2^k(1-\e)^k f(\a,\e)$).}
First, we observe that for all $\e \in [0,1)$,
\begin{eqnarray}
{2^k(1-\e)^k f(\a,\e)} &  = &
2^{-k}{\sum_{j=0}^k\binom{k}{j}(2\a-1)^{j}[(2-\e)^{k-j}\e^{j}-1]^2} \nonumber \\
& \equiv & T_{2}(\a,\e)
+2^{-k}{\sum_{j=2}^k\binom{k}{j}(2\a-1)^{j}[(2-\e)^{k-j}\e^{j}-1]^2}
\nonumber
\\
& \leq & T_{2}(\a,\e) + 2^{-k}{\sum_{j=2}^k\binom{k}{j}(2\a-1)^{j}}  \nonumber \\
& = & T_{2}(\a,\e) + {\a^k -k2^{-k}(2\a-1)-2^{-k}} \enspace .
\label{eq:hard_rest_terms}
\end{eqnarray}

Next, we will prove that
\begin{eqnarray}
{\frac{T_{2}(\a,\e_1)}{2^k(1-\e_1)^k} - f(1/2,\e_0)} & = &
\frac{\left((2-\e_1)^k-1\right)^2}{4^k(1-\e_1)^k} +
\frac{k(2\a-1)\left((2-\e_1)^{k-1}\e_1-1\right)^2}{4^k(1-\e_1)^k}
- \frac{\left((2-\e_0)^k-1\right)^2}{4^k(1-\e_0)^k} \nonumber \\
& < & \a k2^{-2k-1} (1-\e_0)^{-k-1}  \label{eq:first_two} \enspace
.
\end{eqnarray}
For this, define $\Up_1(\e)=1-(2-\e)^{k-1}\e$ so that
$\Up_1(\e_0)=0$. For $\e<\e_0$ we infer that
\begin{equation} \label{Up1}
0< \Up_1(\e) \le 1-(2-\e_0)^{k-1}\e =1-\e/\e_0 \enspace .
\end{equation}
Therefore, the function
$$
\Up_2(\e)=\frac{k(2\a-1)\Up_1(\e)^2}{(1-\e)^k}
$$
satisfies
\begin{equation} \label{Up2}
\Up_2(\e_1) \le \frac{k(2\a-1)}{4(1-\e_1)^k} <
\frac{k(\a-1/2)}{2(1-\e_0)^{k+1}} \enspace .
\end{equation}
Next, define
$$
\Up_3(\e)=\frac{\left((2-\e)^k-1\right)^2}{(1-\e)^k} \enspace .
$$
\dima{Differentiation gives
$$
-\Up_3'(\e) \, = \, k \frac{(2-\e)^k-1}{(1-\e)^{k+1}} \, \Up_1(\e)
\, \leq \,  k \left(\frac{2-\e}{1-\e}\right)^{k}
\frac{\Up_1(\e)}{1-\e_0} \enspace .
$$}
Since $\frac{2-\e}{1-\e}$ is increasing in $\e$, we deduce using
(\ref{Up1}) that for  $\e<\e_0$,
$$
-\Up_3'(\e) \le k \frac{(2-\e_0)^k}{(1-\e_0)^{k+1}}
\left(1-\frac{\e}{\e_0}\right) \,.
$$
As $\int_{\e_1}^{\e_0}\left(1-\e/{\e_0}\right) \, d\e =\e_0/8$, we
conclude that
\begin{equation} \label{Up3}
\Up_3(\e_1)-\Up_3(\e_0) \le  k \frac{(2-\e_0)^k
\e_0}{8(1-\e_0)^{k+1}} \le \frac{k}{4(1-\e_0)^{k+1}} \,.
\end{equation}
Adding the inequalities (\ref{Up2}) and (\ref{Up3}), then dividing
by $4^k$, yields (\ref{eq:first_two}).

Combining~\eqref{eq:hard_rest_terms} and \eqref{eq:first_two} and
requiring $k \geq 6$ for~\eqref{eq:itslate} we get
\begin{eqnarray}
{f(\a,\e_1) - f(1/2,\e_0)}
 & <  &
\frac{T_{2}(\a,\e_1)+\a^k -k2^{-k}(2\a-1)-2^{-k}}{2^k(1-\e_1)^k} -
f(1/2,\e_0)
 \nonumber \\
& <  & \frac{\a k2^{-2k-1}}{(1-\e_0)^{k+1}} + \frac{\a^k
-k2^{-k}(2\a-1)-2^{-k}}{2^k(1-\e_1)^k}   \nonumber\\ & = &
\frac{\a^k-\a
k2^{-k-1}\left(4-\frac{(1-\e_1)^k}{(1-\e_0)^{k+1}}\right)+2^{-k}(k-1)}{2^k(1-\e_1)^k}\nonumber
\\
& < & \frac{\a^k-\a
k2^{-k-1}\left(3-k2^{-k+1}\right)+2^{-k}(k-1)}{2^k(1-\e_1)^k}\label{eq:itslate}
\\
& < & \frac{\a^k-3\a k2^{-k-1}+2^{-k}(k-1)+4^{-k}k^2
}{2^k(1-\e_1)^k} \label{neq:last} \enspace .
\end{eqnarray}
\dima{Observe now that for $k \geq 3$ \eqref{neq:last} establishes
our promised claim $f(\a,\e_1) - f(1/2,\e_0) < 2^{-k+1}$.}
Moreover, combining~\eqref{neq:hard_halflife} and~\eqref{neq:last}
we get~\eqref{eq:opa}, while the fact $\a > 9/10$
implies~\eqref{eq:nice}
\begin{eqnarray}
{\frac{f(1/2,\e_0)}{f(\a,\e_1) - f(1/2,\e_0)}}  & > &(1-\e_1)^k
\times \frac{2^k - 2 -
k2^{-k+1}}{\a^k-3\a k2^{-k-1}+2^{-k}(k-1)+ k^2 4^{-k}} \label{eq:opa} \\
& > & (1-\e_1)^k \times \frac{2^k - 2}{\a^k-3\a
k2^{-k-1}+2^{-k}(k-1)} -(2/3)^k\enspace . \label{eq:nice}
\end{eqnarray}
Recall now that for any $0 < \alpha < 1$ and $0 \le q < \alpha^k$,
\[ \frac{1}{\alpha^k - q}  \ge  1 + k(1-\alpha) + q \enspace . \]
Observe that $3\a k2^{-k-1}-2^{-k}(k-1) < \a^k$ for $\a \geq 2/3$.
Since $\a > 9/10$, we thus have
$$
\frac{1}{\a^k-3\a k2^{-k-1}+2^{-k}(k-1)} \ge  1+k(1-\a)+3\a
k2^{-k-1}-2^{-k}(k-1)\enspace .
$$

Thus, by~\eqref{eq:taste},\eqref{neq:hard_halflife}
and~\eqref{eq:nice} we see that~\eqref{neq:hard_est} holds as long
as $r < (1-\e_1)^k\phi(\a)-2\times(2/3)^k$ where
\[
 \phi(\alpha) \equiv \bigl(\log 2 - h(\alpha) \bigr)
 \left( 2^{k}(k+1)-3k-\frac{1}{2} -\a k\left(2^k-\frac{7}{2}\right)\right)
  .
\]

We are thus left to minimize $\phi$ in $(9/10,1]$. It will be
convenient to define
\begin{eqnarray}
B & = & 2^k(k+1)-3k -\frac{1}{2}\\
C & = & k\left(2^k-\frac{7}{2}\right)
\end{eqnarray}
and rewrite
$$
\phi (\a)=\bigl(\log 2 - h(\alpha) \bigr)  \left(B-\alpha C\right)
\enspace .
$$
Since $\phi$ is {differentiable} its minima can only occur at
$9/10,1$ or where
\begin{equation} \label{eq:hard_phiprime} \phi'(\a) = \log\left(\frac{\a}{1-\a}\right)\left(B-\a
C\right)-\left(\log 2 - h(\a)\right) C  = 0 \enspace .
\end{equation}
Note now that
\[ \lim_{\alpha \to 1}  \frac{\phi'(\a)}{\log(1-\a)} = -(B-C) < 0\] and, thus, the
derivative of $\phi$ becomes positively infinite as we approach 1.
At the same time,
\[ \phi'(9/10) < 2.2B-2.3C \]
which is negative for $k \ge 23$.    Therefore, $\phi$ is
minimized in the interior of $(9/10,1]$ for all $k\geq 23$.
Setting the derivative of $\phi$ to zero gives
\begin{align}
- \log (1-\alpha) & =  (\log 2 - h(\a)) \times \frac{C}{B-\a C} - \log \a \nonumber \\
 & =  (\log 2 - h(\alpha)) \times \frac{k}
        {1 + k (1-\alpha) +\frac{k+6}{2^{k+1}-7}}
 - \log \alpha \label{eq:boot} \enspace .
\end{align}

By ``bootstrapping'' we will derive a tightening series of bounds
on the solution of~\eqref{eq:boot} in $\alpha \in (9/10,1)$.  Note
first that we have an easy upper bound,
\begin{equation}
\label{eq:phiupper} - \log (1-\alpha) < k \log 2 - \log \alpha
\enspace .
\end{equation}
At the same time, if $k \geq 3$ then $(k+6)/(2^{k+1}-7) \leq 1$,
implying
\begin{equation}
\label{eq:philower}
 - \log (1-\alpha)
 \geq \frac{k \,(\log 2 - h(\alpha))}
        {2 + k (1-\alpha)}
 - \log \alpha
\enspace .
\end{equation}
If we write $k(1-\alpha)=\dima{D}$ then~\eqref{eq:philower}
becomes
\begin{equation}\label{eq:contra}
- \log (1-\alpha)
 \geq  \frac{\log 2 - h(\alpha)}{1-\alpha}
  \left( \frac{D}{D+2} \right)
  - \log \alpha \enspace .
\end{equation}

By inspection, if $D \geq 3$ the r.h.s.\ of~\eqref{eq:contra} is
greater than the l.h.s.\ for all $\alpha > 9/10$, yielding a
contradiction. Therefore, $k(1-\alpha) < 3$ for all $k \geq 3$.
Since $\log 2 - h(\alpha) > 0.36$ for $\alpha > 9/10$, we see that
for $k\geq 3$, \eqref{eq:philower} implies
\begin{eqnarray}
- \log (1-\alpha) & > & 0.07 \,k \label{eq:rissoto}
\qquad\mbox{or,
equivalently,}\\
1-\a & < & \ee^{-0.07 \, k} \enspace . \label{eq:rissota}
\end{eqnarray}
Observe now that \eqref{eq:rissota} implies
\begin{equation}\label{eq:lotus}
k (1-\alpha) < k \, \ee^{-0.07 k} \enspace ,
\end{equation}
and, hence, as $k$ increases the denominator of~\eqref{eq:boot}
actually approaches $1$.

To bootstrap, we first note that since $\alpha
> 1/2$ we have
\begin{eqnarray}
 h(\alpha) & \le & -2 (1-\alpha) \log (1-\alpha) \label{eq:ent_b_l}\\
           & < & 2 \,\ee^{-0.07 \,k} (k \log 2 - \log 0.9) \label{eq:randkl} \\
           & < & 2 \,k \,\ee^{-0.07 \,k} \label{eq:more_ent}
\end{eqnarray}
where~\eqref{eq:randkl} relies on~\eqref{eq:rissota} and~
\eqref{eq:phiupper}. Moreover, $\alpha > 1/2$ implies $-\log
\alpha \le 2(1-\alpha)$ which, by~\eqref{eq:rissota} implies
$-\log \alpha < 2 \,\ee^{-0.07 \,k}$. Thus, starting
with~\eqref{eq:boot}, using~\eqref{eq:lotus}, taking $k \geq 3$
and using~\eqref{eq:more_ent}, and finally using $1/(1+x)>1-x$ for
all $x
> 0$ we get
\begin{eqnarray}
 -\log (1-\alpha) & > & \frac{k \,(\log 2 - h(\alpha))}
                            {1 + k \,\ee^{-0.07 \,k} + \dima{\frac{k+6}{2^{k+1}-7}}} \nonumber \\
  & > & \frac{k \,(\log 2 - 2 \,k \,\ee^{-0.07 \,k}) }
             {1 + 2 \,k \,\ee^{-0.07 \,k}} \nonumber \\
  & > & k \,(\log 2 - 2 \,k\,\ee^{-0.07 \,k}) (1 - 2 \,k \,\ee^{-0.07 \,k} )
\nonumber \\
  & > & k \log 2 - 4 \,k^2\,\ee^{-0.07 \,k} \enspace . \label{eq:fried}
\end{eqnarray}
For $k \ge 166$, $4 \,k^2 \,\ee^{-0.07 \,k} < 1$. Thus, for such
$k$, \eqref{eq:fried} implies $1 - \alpha < 3 \times 2^{-k}$.
This, in turn, implies $-\log \alpha \leq 2(1-\alpha) < 6 \times
2^{-k}$ and so, by \eqref{eq:ent_b_l} and~\eqref{eq:phiupper}, we
have that for all $k \geq 166$ and $\alpha > 9/10$
\begin{equation}\label{eq:entr_last}
h(\alpha) < 6 \times 2^{-k} (k \log 2 - \log \alpha)
          < 5 \,k \,2^{-k} \enspace .
\end{equation}

Plugging~\eqref{eq:entr_last} into~\eqref{eq:boot} to bootstrap
again, we get (analogously to the derivation of\eqref{eq:fried})
that
\begin{eqnarray*}
 -\log (1-\alpha) & > & \frac{k \,(\log 2 - 5 \,k \,2^{-k})}
                            {1 + 3 \,k\,2^{-k} + \dima{\frac{k+6}{2^{k+1}-7}}} \\
 & > &  \frac{k \,(\log 2 - 5 \,k \,2^{-k})}
             {1 + 6 \,k \,2^{-k}}\\
 & > & k \,(\log 2 - 5 \,k \,2^{-k}) (1 - 6 \,k \,2^{-k})\\
  & > & k \log 2 - 11 \,k^2 \,2^{-k}  \enspace .
\end{eqnarray*}
Since $\ee^x < 1+2x$ for $x < 1$ and $11 \,k^2 \,2^{-k} < 1$ for
$k > 10$, we see that
\[ 1-\alpha < 2^{-k} + 22 \,k^2 \,2^{-2k} \enspace .
\]

Plugging into~\eqref{eq:phiupper} the fact $-\log \alpha < 6
\times 2^{-k}$ we get $-\log (1-\alpha) < k \log 2 + 6 \times
2^{-k}$. Using that $\ee^{-x} \geq 1-x$ for $x\geq 0$, we get the
closely matching upper bound,
\[ 1-\alpha > 2^{-k} - 6 \times 2^{-2k} \enspace . \]

Thus, we see that for $k \ge 166$, $\phi$ is minimized at an
$\alpha_{\min}$ which is within $\delta$ of $1-2^{-k}$, where
$\delta = 22 \,k^2 \,2^{-2k}$. Let $T$ be the interval
$[1-2^{-k}-\delta, 1-2^{-k}+\delta]$. Clearly the minimum of
$\phi$ is at least
$$
\phi(1-2^{-k}) - \delta \times \max_{\alpha \in T}
\left|\phi'(\a)\right| \enspace . $$ \dima{Using very crude
bounds,} it is easy to see from~\eqref{eq:hard_phiprime} that if
$\alpha \in T$ then $|\phi'(\a)| \le 2 \,k \,2^k$.

Now, since for $k \ge 1$ we have $\log (1-2^{-k}) > - 2^{-k} -
2^{-2k}$, a simple calculation gives
\begin{equation}
 \phi(1-2^{-k}) >  2^k \log 2 + \frac{\log 2}{2} (k - 1) -1 -2 k^2 \, 2^{-k}
\enspace .
\end{equation}
Therefore,
\[ \phi_{\min}
 >  2^k \log 2 + \frac{\log 2}{2} (k - 1)- 1 - 46 \,k^3 \,2^{-k}
\enspace .
\]
Finally, recall that~\eqref{neq:hard_est} holds as long as $r <
(1-\e_1)^k\phi_{\min}-2\times(2/3)^k$. Using the upper bound for
$\e_0$ from~\eqref{eq:eps_bound} we get
\begin{eqnarray*}
{\left(1-\e_1\right)^k \times \phi_{\min}}  & > &\left(1-2^{-k} -
\frac{2k}{4^{k}}\right)^k \!\times \left(2^k
\log 2 + \frac{\log 2}{2} (k - 1)-1- \frac{46 k^3}{2^{k}}\right ) \\
& > &  \left(1-k 2^{-k} - \frac{2k^2}{4^{k}}\right) \!\times\!
\left(2^k \log 2
+ \frac{\log 2}{2} (k - 1) -1 - \frac{46 k^3}{2^{k}}\right )\\
& > & 2^k \log 2 - \frac{\log 2}{2} (k+1) - 1 - \frac{50 k^3}{2^{k}}\\
& = & \rho_k \enspace .
\end{eqnarray*}

\section{Bounds for specific values of  $k$}\label{sec:concrete}

Recall from our discussion in Section~\ref{ground} that in order
to establish $r \geq r_k$ it suffices to prove that there exists
some $\e \in [0,1)$ for which the function $g_r$ defined
in~\eqref{eq:gr_def}, \ie
\begin{equation}
  g_r(\alpha) =
    \frac{f(\alpha)^r}{\alpha^{\alpha}(1-\alpha)^{1-\alpha}}
    =  \frac{\left((2-2\e+\a\e^2)^k - 2
    (1-\e+\a\e)^k+\a^k\right)^r}{\alpha^{\alpha}(1-\alpha)^{1-\alpha}}
    \enspace ,
    \label{eq:to_plot}
\end{equation}
has a unique global maximum at 1/2. Recall also that  for any $r$
the only choice of $\e$ for which $g''_r(1/2)<0$ is the one
mandated by~\eqref{eq:studly}. Thus, for any fixed $k$ one can get
a lower bound for $r_k$ by: i) solving~\eqref{eq:studly}, ii)
substituting the solution to~\eqref{eq:to_plot}, and iii) plotting
the resulting function to check whether $g_r(1/2)>g_r(\alpha)$ for
all $\alpha \neq 1/2$. As $g_r$ never has more than three local
maxima this is very straightforward and yields the lower bounds
referred to as ``simple" lower bounds in Table~\ref{tab:naiv_val}
below.

As mentioned in the Introduction, the simple weighting scheme
yielding Theorem~\ref{easy} does not yield the best possible lower
bound afforded by applying the second moment method to balanced
satisfying assignments. For that, one has to use the significantly
more refined argument which we presented in
Sections~\ref{trunc}---\ref{end_trunc}. That argument also
eventually reduces to proving $g_r(1/2)>g_r(\a)$ for all $\a \neq
1/2$. Now, though, $\e$ is allowed to depend on $\alpha$, subject
only to $\e \leq \e_0$, where $\e_0$ is the solution
of~\eqref{eq:studly}. Naturally, at $\alpha=1/2$ one still has to
take $\e = \e_0$ so that the derivative of $g_r$ vanishes, but for
larger $\a$ (where the danger is) it turns out that decreasing
$\e$ somewhat helps. The bounds reported in Table~\ref{tab:val} in
the Introduction (and replicated below as the ``refined" bounds)
are, indeed, the result of such optimization of $\e$ as a function
of $\a$.

Specifically, for $k \leq 5$ we considered 10,000 equally spaced
values of $\alpha \in [0,1]$ and for each such value found $\e
\leq \e_0$ such that the condition $g_r(\alpha,\e) <
g_r(1/2,\e_0)$ holds with a bit of room. (For $k>4$ we
solved~\eqref{eq:studly}, defining $\e_0$, numerically to 10
digits of accuracy. For the optimization we exploited convexity to
speed up the search.) Having determined such values of $\e$, we
(implicitly) assigned to every not-chosen point in $[0,1]$ the
value of $\e$ at the nearest chosen point. Finally, we computed a
(crude) upper bound on the derivative of $g_r$ with respect to
$\alpha$ in $[0,1]$. This bound on the derivative, along with our
room factor, then implied that for every point that we did not
check, the value of $g_r$ was sufficiently close to its value at
the corresponding chosen point to also be dominated by
$g_{r}(1/2,\e_0)$. For $k>5$, we only partitioned $[0,1]$ into two
intervals, namely $[1/10,9/10]$ and its complement. Assigning the
values  $\e_0$ and $\e_0/2$, respectively, to all the points in
each interval yielded the bounds for such $k$.

\begin{table}[h]\label{tab:naiv_val}
\centering $
\begin{array}{c|ccccccc}
    k                       &  3     & 4     & 5      & 7      & 10      & 20       & 21 \\
    \hline             \\
    \mbox{Upper bound}      & 4.51   & 10.23 & 21.33  & 87.88  & 708.94  & 726,817  & 1,453,635 \\
    \mbox{Refined lower bound}    & 2.68   & 7.91  & 18.79  & 84.82  & 704.94  & 726,809  & 1,453,626 \\
    \mbox{Simple  lower bound}      & 2.54   & 7.31  & 17.61  & 82.63  & 701.53  & 726,802  & 1,453,619 \\
\end{array}
$
\end{table}

\section{Conclusions}\label{sec:conc}

\dima{We proved that the random $k$-SAT threshold satisfies $r_k
\sim 2^k \log 2$. In particular, we proved that random $k$-CNF
formulas with density $2^k \log2 - k (\log 2)/2-O(1)$ have
exponentially many balanced satisfying truth assignments. That is,
truth assignments that have at least one satisfied literal in
every clause yet, in total, satisfy only as many literal
occurrences as a random truth assignment.

Our argument leaves a gap of order $\Theta(k)$ with the first
moment upper bound $2^k \log 2$.} With respect to this gap it is
worth pointing out that the best known techniques~\cite{DuBo,KKKS}
for improving this upper bound only give $r_k \leq 2^k \log 2 -
b_k$ where $b_k \to (1+\log 2)/2$.  At the same time, it is not
hard to prove that for $r = 2^k \log 2 - k(\log 2)/2$, \ie within
an additive constant from our lower bound, \whp\ there are no
satisfying truth assignments that satisfy only $km/2+o(km)$
literal occurrences. Thus, any asymptotic improvement over our
lower bound would mean that tendencies toward the majority
assignment become essential as we approach the threshold.

The gap between the upper bound and the best algorithmic lower
bound $r_k=\Omega(2^k/k)$, seems \dima{to us} much more
significant (and is certainly much bigger!). The lack of progress
in the last ten years suggests the possibility that no polynomial
time algorithm can improve the lower bound asymptotically. {At the
same time, in a completely different direction, M\'{e}zard and
Zecchina~\cite{MZ} recently used the \dima{non-rigorous} cavity
method of statistical physics to obtain detailed predictions for
the satisfiability threshold suggesting that $r_k=2^k \log 2
-O(1)$. (See also \cite{MPZ} for an overview.) Insights from this
analysis led them to an intriguing algorithm called ``survey
propagation'' (described in \cite{MZ,BMZ}) that seems to perform
well on random instances of $k$-SAT close to the threshold, at
least for small $k$. (Its performance is especially impressive for
$k=3$.) A rigorous analysis of this algorithm is \dima{still
lacking, though, and it remains unclear whether its success for
values of $r$ close to the threshold extends to large $k$.}}

 The
success of the second moment method for balanced satisfying truth
assignments suggests that such assignments form a ``mist'' in
$\{0,1\}^n$ and, as a result, they might be hard to find by
algorithms based on local updates. Moreover, as $k$ increases the
influence exerted by the majority vote assignment becomes less and
less significant as most literals occur very close to their
expected $kr/2$ times. As a result, the structure of the space of
solutions may well be different for small $k$ (e.g. $k=3,4$) and
for larger $k$.\medskip

 To summarize, the following key questions remain:
\begin{enumerate}
\item Is $2^k\log 2 -r_k$ bounded?
\item
Is there an {\bf algorithmic threshold} $\lambda_k=o(2^k)$ so that
for $r>\lambda_k$, no polynomial-time algorithm can find a
satisfying truth assignment for the random formula $F_k(n,rn)$
with uniformly positive probability?
\end{enumerate}

\section*{Acknowledgments}
We are grateful to Cris Moore for illuminating conversations and
to Mike Molloy for helpful comments. \dims{We are indebted to
Chris Calabro for a careful reading and for a correction to a
previous version of this paper.} Part of this work was done while
the authors participated in the focused research group on discrete
probability at BIRS, July 12-26, 2003.

 \def\bibRCS{$Id: theory.bib,v 1.359 2000/09/08 01:00:19 beame Exp beame $}
  \makeatletter \@ifundefined{ccisdefined}{ \newcommand{\cc}[1]{\mbox{\it
  #1\/}} \newcommand{\ccisdefined}{} }{} \@ifundefined{journalfont}{
  \newcommand{\journalfont}{\it } }{} \makeatother \def\bibRCS{$Id:
  theory.bib,v 1.366 2000/09/15 17:54:09 beame Exp $} \makeatletter
  \@ifundefined{ccisdefined}{ \newcommand{\cc}[1]{\mbox{\it #1\/}}
  \newcommand{\ccisdefined}{} }{} \@ifundefined{journalfont}{
  \newcommand{\journalfont}{\it } }{} \makeatother


\begin{thebibliography}{10}

\bibitem{naesat}
D.~Achlioptas and C.~Moore.
\newblock The asymptotic order of the random $k$-{SAT} threshold.
\newblock In {\em Proc. 43th Annual Symposium on Foundations of Computer Science},
pages 126--127, 2002.




\bibitem{BMZ}
A.~Braunstein, M.~M\'{e}zard,  and R.~Zecchina.
\newblock Survey propagation: an algorithm for satisfiability.
\newblock Preprint, 2002.

\bibitem{ChFrUC}
M.-T. Chao and J.~Franco.
\newblock Probabilistic analysis of a generalization of the unit-clause literal
  selection heuristics for the $k$-satisfiability problem.
\newblock {\em Inform. Sci.}, 51(3):289--314, 1990.


\bibitem{cheese}
P.~Cheeseman, B.~Kanefsky, and W.~Taylor.
\newblock Where the really hard problems are.
\newblock In {\em Proc. 12th International Joint Conference on Artificial
  Intelligence (IJCAI-91) Vol. 1}, pages 331--337, 1991.

\bibitem{mick}
V.~Chv\'{a}tal and B.~Reed.
\newblock Mick gets some (the odds are on his side).
\newblock In {\em Proc. 33th Annual Symposium on Foundations of Computer
  Science}, pages 620--627, 1992.

\bibitem{debruijn}
N.~G. de~Bruijn.
\newblock {\em Asymptotic methods in analysis}.
\newblock Dover Publications Inc., New York, 3rd edition, 1981.


\bibitem{DPRZ} A. Dembo, Y. Peres, J. Rosen and O.
Zeitouni.
\newblock Thick points for planar Brownian motion
and the Erd\H{o}s-Taylor conjecture on random walk.
\newblock {\em Acta Math.}, 186:239--270, 2001.


\bibitem{dez}
A. Dembo and O. Zeitouni.
\newblock {\em Large deviations techniques and applications}.
\newblock Springer Verlag, New York, 2nd edition, 1998.


\bibitem{DuBo}
O.~Dubois and Y.~Boufkhad.
\newblock A general upper bound for the satisfiability threshold of random
  $r$-{S}{A}{T} formulae.
\newblock {\em J. Algorithms}, 24(2):395--420, 1997.


\bibitem{dub_announ}
O.~Dubois, Y.~Boufkhad, and J.~Mandler.
\newblock Typical random {3-SAT} formulae and the satisfiability threshold.
\newblock In {\em Proc. 11th Annual Symposium on Discrete Algorithms},
  pages 126--127, 2000.


\bibitem{LLL}
P. Erd{\H{o}}s and L.~Lov{\'a}sz.
\newblock Problems and results on $3$-chromatic hypergraphs and some related questions.
\newblock Colloq. Math. Soc. J\'anos Bolyai, Vol. 10, 609--627.


\bibitem{ET}
P. Erd\H{o}s and S. J. Taylor.
\newblock {Some problems concerning the structure
of random walk paths}.
\newblock  {\em Acta Sci. Hung.} 11:137--162, 1960.


\bibitem{FrPa}
J.~Franco and M.~Paull.
\newblock Probabilistic analysis of the {Davis--P}utnam procedure for solving
  the satisfiability problem.
\newblock {\em Discrete Appl. Math.}, 5(1):77--87, 1983.

\bibitem{frie}
E.~Friedgut.
\newblock Necessary and sufficient conditions for sharp thresholds of graph
  properties, and the {$k$-SAT} problem.
\newblock {\em J. Amer. Math. Soc.}, 12:1017--1054, 1999.


\bibitem{FrSu}
A.~M. Frieze and S.~Suen.
\newblock Analysis of two simple heuristics on a random instance of
  $k$-{S}{A}{T}.
\newblock {\em J. Algorithms}, 20(2):312--355, 1996.

\bibitem{friezewormald}
A.~Frieze and N.~C. Wormald.
\newblock Random {$k$-SAT}: a tight threshold for moderately growing $k$.
\newblock In {\em Proc. 5th International Symposium on Theory and
  Applications of Satisfiability Testing}, pages 1--6, 2002.

\bibitem{JSV}
S.~Janson, Y.~C. Stamatiou, and M.~Vamvakari.
\newblock Bounding the unsatisfiability threshold of random {3-SAT}.
\newblock {\em Random Structures Algorithms}, 17(2):103--116, 2000.

\bibitem{342}
A.~Kaporis, L.~M. Kirousis, and E.~G. Lalas.
\newblock The probabilistic analysis of a greedy satisfiability algorithm.
\newblock In {\em Proc. 10th Annual European Symposium on Algorithms},
volume 2461 of {\em Lecture Notes in Computer Science}, pages
574--585. Springer, 2002.

\bibitem{KKKS}
L.~M. Kirousis, E.~Kranakis, D.~Krizanc, and Y.~Stamatiou.
\newblock Approximating the unsatisfiability threshold of random formulas.
\newblock {\em Random Structures Algorithms}, 12(3):253--269, 1998.

\bibitem{MPZ}
M.~M\'{e}zard, G. Parisi, and R.~Zecchina.
\newblock Analytic and Algorithmic Solution of Random
Satisfiability Problems.
\newblock {\em Science}, 297: 812-815, 2002.

\bibitem{MZ}
M.~M\'{e}zard and R.~Zecchina.
\newblock Random {$K$}-satisfiability: from an analytic solution to a new efficient
algorithm.
\newblock {\em Phys. Rev. E}, 66, 056126, 2002.

\bibitem{MSL}
D.~G. Mitchell, B.~Selman, and H.~J. Levesque.
\newblock Hard and easy distributions of {SAT} problems.
\newblock In {\em Proc. 10th National Conference on Artificial Intelligence},
  pages 459--462, 1992.

\bibitem{MZRK}
R.~Monasson and R.~Zecchina.
\newblock Statistical mechanics of the random ${K}$-satisfiability model.
\newblock {\em Phys. Rev. E (3)}, 56(2):1357--1370, 1997.

\bibitem{drunk}
I. Stewart.
\newblock Where drunkards hang out.
\newblock {\em Nature}, News and Views,
October 18, 2001.

\end{thebibliography}
\end{document}